\documentclass[onecolumn,twoside]{IEEEtran}
\usepackage{mathpazo}
\usepackage{times}
\usepackage{amsmath}
\usepackage{amsfonts}
\usepackage{latexsym}
\usepackage{amssymb}
\usepackage{mathrsfs}

\usepackage{upref}
\usepackage{theorem}
\usepackage{graphicx}
\usepackage{psfrag}
\usepackage{cite}

\usepackage{color}

\theoremstyle{plain}
\theorembodyfont{\normalfont\slshape}

\newtheorem{thm}{Theorem$\!$}
\newenvironment{theorem}
{\begin{thm}\hspace*{-1ex}{\bf.}}{\end{thm}}

\newtheorem{lem}[thm]{Lemma$\!$}
\newenvironment{lemma}{\begin{lem}\hspace*{-1ex}{\bf.}}{\end{lem}}

\newtheorem{alg}[thm]{Algorithm$\!$}
\newenvironment{algorithm}{\begin{alg}\hspace*{-1ex}{\bf.}}{\end{alg}}

\newtheorem{prop}[thm]{Proposition$\!$}

\newtheorem{cor}[thm]{Corollary$\!$}
\newenvironment{corollary}{\begin{cor}\hspace*{-1ex}{\bf.}}{\end{cor}}

\newtheorem{defn}[thm]{Definition$\!$}
\newenvironment{definition}{\begin{defn}\hspace*{-1ex}{\bf.}}{\end{defn}}

\newtheorem{xmpl}[thm]{Example$\!$}
\newenvironment{example}{\begin{xmpl}\hspace*{-1ex}{\bf.}}{\hfill$\Box$\end{xmpl}}

\newtheorem{cnstr}{Construction$\!$}

\newcounter{enumrom}
\renewcommand{\theenumrom}{(\roman{enumrom})}

\makeatletter
\renewcommand{\@endtheorem}{\endtrivlist}
\makeatother

\makeatletter
\renewcommand{\thefigure}{{\@arabic\c@figure}}
\renewcommand{\fnum@figure}{{\bf Figure\,\thefigure}}
\makeatother

\newcommand{\cV}{\mathcal{V}}

\newcommand{\pf}{{\bf Proof: }}

\newcommand{\be}[1]{\begin{equation}\label{#1}}
\newcommand{\ee}{\end{equation}}

\renewcommand{\leq}{\leqslant}

\renewcommand{\geq}{\geqslant}

\newcommand{\Cref}[1]{Co\-ro\-lla\-ry\,\ref{#1}}

\newcommand{\C}{\mbox{${\cal C}$}}

\newcommand{\qed}{\hfill$\Box$\\[1ex]}
\newcommand{\hs}{\mbox{$\hat{s}$}}
\newcommand{\uc}{\mbox{$\underline{c}$}}
\newcommand{\uu}{\mbox{$\underline{u}$}}
\newcommand{\al}{\alpha}
\newcommand{\uzero}{\underline{0}}
\newcommand{\xor}{\oplus}

\newcommand{\eq}{\mbox{$\,=\,$}}

\outer\def\proclaim #1. #2\par{\medbreak
 \noindent{\bf#1.\enspace}{\sl#2\par}%
 \ifdim\lastskip<\medskipamount \removelastskip\penalty55\medskip\fi}

\begin{document}


\title{\LARGE\bf Extended Integrated Interleaved Codes over any Field
with Applications to Locally Recoverable Codes}

\author{\large
Mario~Blaum \\
IBM Research Division\\
Almaden Research Center\\
 San Jose, CA 95120, USA \\
Mario.Blaum@ibm.com}


\maketitle

\begin{abstract}
Integrated Interleaved (II) and Extended Integrated Interleaved (EII)
codes are a versatile alternative for Locally Recoverable (LRC)
codes, since they require fields of relatively small size.
II and EII codes are generally defined over Reed-Solomon type of codes. A new
comprehensive definition of EII codes is presented,
allowing for EII codes over any field, and in particular, over
the binary field $GF(2)$. The traditional definition
of II and EII codes is shown to be a special case of the new
definition.  Improvements over previous constructions of LRC codes,
in particular, for binary codes, are given, as well as cases meeting
an upper bound on the minimum distance.
Properties of the codes are presented as well, in particular,
an iterative decoding algorithm on rows and columns generalizing the
iterative decoding algorithm of product codes. Two applications are
also discussed: one is finding a systematic encoding of EII
codes such that the parity symbols have a balanced distribution on
rows, and the other is the problem of
ordering the symbols of an EII code such that the maximum length of a
correctable burst is achieved.
\end{abstract}


\section{Introduction}
\label{Introduction}

Normally, error and erasure correcting codes are studied at an
individual level~\cite{ms}. Codewords are transmitted or stored
independently, one after the other. However, there is a vast
literature on codes in which a set of codewords (that in general we
will call an array) are connected with each other through some extra
parities affecting all of the individual codewords. That way, the
array has extra protection and when the error-erasure correcting
capability of one of the codewords is exceeded, often the extra
parities allow for correction of such an event. Examples of codes
with these characteristics are tensor and generalized tensor
codes~\cite{hys,hys2,if,wolf}, generalized concatenated codes~\cite{bz,z},
pyramid codes~\cite{hcl}, PMDS codes (also known as maximally
recoverable codes)~\cite{bhh,bpsy,gybs,ggy,ghjy,hy,hsx}, sector-disk codes
(SD)~\cite{bpsy,hy,ld,pb,pbh}, locally recoverable
codes (LRC)~\cite{bpk,bkv,bh,ghsy,hx,hxc,kbty,kna,lxy,pd,pklk,rk,rp,sa,scyh,sd,sy,tb,wz},
multilevel codes~\cite{pa,wc}, stair
codes~\cite{ll,ll2} and integrated and extended integrated interleaved (II
and EII) codes~\cite{bh,bh2,hapkt,tk,w,z2,z3}. Certainly, this list is
not complete and several results are common to different
approaches. There is also recent work on LRC codes with the rank
metric~\cite{keds}.

We will consider the framework of LRC codes for erasure correction.
In general, in this type of codes the
data symbols are divided into sets and parity symbols
(i.e., local parities) are added to each set (often, but not
necessarily, using an MDS code).
This way, when a number of erasures smaller than the minimum distance
occurs in a set, such erasures are rapidly
recovered. In addition to the local parities, a number of
global parities are added. Those
global parities involve all of the
data symbols and may include the local parity symbols as well. The
goal of the extra parities, as stated above, is to correct situations in which the
erasure-correcting power of the local parities has been exceeded.

Since the codes we are studying can be viewed as special cases of LRC
codes, let us state their definition formally, which is similar to
the definition of multi-erasure locally recoverable codes (ME-LRC)
given in~\cite{hys}.

\begin{definition}
\label{defLRC}
{\em
Consider a code $\C$ over a finite field $GF(q)$ consisting of
$m\times n$ arrays such that, given integers $h$ and $g$
where $1\leq h\,<\,n$ and $0\leq g\,<\,m(n-h)$, the arrays
satisfy:

\begin{enumerate}

\item Each row in each array in $\C$ is in an
$[n,n-h,d_0]$ code over $GF(q)$.

\item Reading the symbols of $\C$ row-wise, $\C$ is an
$[mn,k,d]$ code over $GF(q)$, where $k\eq m(n-h)-g$.

\end{enumerate}

Then we say that $\C$ is an $(m,n,k;h,d_0,d,q)$ LRC code.\qed
}
\end{definition}

There are well known bounds for the minimum distance $d$ of an
$(m,n,k;h,d_0,d,q)$ LRC code. A Singleton type of bound is provided
in~\cite{ghsy,pklk}. An expression of the bound in tune with the
notation above is given in~\cite{bh,bh2} as follows:

\begin{eqnarray}
\label{d0hg}
d&\leq &\left\lceil {\frac {g+1}{n-h}} \right\rceil h\,+\,g\,+\,1.
\end{eqnarray}

Notice that, if $m\eq 1$, all the parities are local, i.e., $g\eq 0$
and bound~(\ref{d0hg}) is the usual Singleton bound~\cite{ms}. With
some exceptions~\cite{bh,bh2}, the best constructions of codes
meeting bound~(\ref{d0hg}) require a field $GF(q)$ of size at least
$mn$~\cite{tb}. Bound~(\ref{d0hg}), being a Singleton type of bound,
does not take into account the size $q$ of the field. Bounds
considering the size $q$ are provided in~\cite{cm,cm2}. Using the
formulation of this bound given in~\cite{hys}, we have

\begin{eqnarray}
\label{boundcmd}
d&\leq &\min\left\{d_{\rm opt}^{(q)}\,[(m-j)n\,,\,k-jk^*]\quad {\rm
for}\quad 0\leq j\leq \left\lceil\frac{k}{k^*}\right\rceil-1\right\},
\end{eqnarray}
where 
$d_{\rm opt}^{(q)}\,[N,K]$ denotes the largest
possible minimum distance of a linear code of length $N$ and
dimension $K$ over $GF(q)$, $k_{\rm opt}^{(q)}\,[N,D]$ the
largest dimension $K$ of a linear code of length $N$ and minimum
distance $D$, and $k^*\eq k_{\rm opt}^{(q)}\,[n,d_0]$. Some recent
papers~\cite{ab,wzl} have improved upon this bound.

In Definition~\ref{defLRC}, each row corresponds to a parity set.
Strictly speaking, it is not necessary that the
parity sets as given by rows in this description are disjoint. For
example, Definition~1 of local-error correction (LEC) codes
in~\cite{pklk} does not
make this assumption; however, most constructions do, like for
example, those given in~\cite{tb},

Let us point out that the interest in erasure correcting codes with
local and global properties arises mainly from
two applications. One of them is the cloud. A cloud configuration
may consist of many storage devices, of which some of them may even
be in different geographical locations, and the data is distributed
across them. If one or more of those devices fails, it is
desirable to recover its contents ``locally,'' that is, using a few
parity devices within a set of limited size in order to affect
performance as little as possible. However, the local parities may not
suffice. Extra protection is needed in case the erasure-correcting
capability of a local set is
exceeded. To address this
situation, some devices consisting of global parities are
incorporated. When the local correction power is exceeded,
the global parity devices
are invoked and correction is attempted. If such a situation occurs,
there will be an impact on performance, but
data loss may be averted. It is expected that the cases in which
the local parity is
exceeded are relatively rare events, so the aforementioned impact on
performance does not occur frequently.
As an example of this type of application, we refer the reader to the
description of the Azure system~\cite{hsx} or to the Xorbas code
presented in~\cite{sa}.

A second application occurs in the context of Redundant Arrays of
Independent Disks (RAID) architectures~\cite{g}. In this case, a RAID
architecture protects against one or more storage device failures.
For example, RAID 5 adds one extra parity device, allowing for the
recovery of the contents of one failed device, while RAID 6 protects
against up to two device failures. In particular, if those devices
are Solid State Drives (SSDs), like flash memories, their
reliability decays with time and with the number of writes and
reads.
The information in SSDs is generally divided into pages, each page
containing its own internal Error-Correction Code (ECC). It may
happen that a particular page degrades and its ECC is exceeded.
However, the user may not become aware of this situation until the page is
accessed (what is known as a silent failure). Assuming an SSD has
failed in a RAID~5 scheme, if during reconstruction a silent page
failure is encountered in one of the surviving SSDs, then data loss
will occur. A method around this situation is using RAID~6. However,
this method is costly, since it requires two whole SSDs as parity. It is more
desirable to divide the information in a RAID type of architecture
into $m\times n$ stripes: $m$ represents the size of a stripe, and
$n$ is the number of SSDs. The RAID architecture may be viewed as
consisting of a large number of stripes, each stripe encoded and
decoded independently. Certainly, codes like the ones used in cloud
applications may be used as well for RAID applications. In practice,
the choice of code depends on the
statistics of errors and on the frequency of silent page failures.
RAID systems, however, may behave differently than a cloud array of devices, in
the sense that each column represents a whole storage device. When a device
fails, then the whole column is lost, a correlation that may not
occur in cloud applications. For that reason, RAID architectures may
benefit from a special class of codes with local and global
properties, the so called sector-disk (SD)
codes, which take into
account such correlations~\cite{hy,ld,pb,pbh}.

From now on, we will call the entries of the codes considered in the
paper ``symbols''. Such symbols can be whole devices (for example, in
the case of cloud applications) or pages (in the case of RAID
applications for SSDs). Each symbol may be protected
by one local group, but a natural extension is to consider multiple
localities~\cite{rp,tb,zy}. Product codes~\cite{ms} represent a
special case of multiple localities: any symbol is protected by both
horizontal and vertical parities.

Product codes by themselves may also be used in RAID-type of
architectures: the horizontal parities protect a number of devices
from failure. The vertical parities allow for rapid recovery of a page
or sector within a device (a first responder type of approach).
However, if the number of silent failures exceeds the correcting
capability of the vertical code, and the horizontal code is unusable
due to device failure, data loss will occur. For that reason, it may be
convenient to incorporate a number of extra global parities to the product
code. We refer to these extra global parities simply as extra parities
in order to avoid confusion, since in a product code
the parities on parities, by affecting all of the symbols, can also
be regarded as global parities.

In effect, consider a product code consisting of $m\times
n$ arrays such that each column has $v$ parity symbols and each row
has $h$ parity symbols. If in addition to the horizontal and vertical
parities there are $g$ extra parities, we say that the code is an
Extended Product (EPC) code and we denote it by $EP(m,v;n,h;g)$.
Notice that, in particular, $EP(m,v;n,h;0)$ is a regular product
code, while $EP(m,0;n,h;g)$ is a Locally Recoverable (LRC)
code~\cite{ghjy,tb}. Constructions and bounds for EPC codes were
presented in~\cite{bh2}.

Constructions of LRC codes involve different issues and tradeoffs,
like the size of the field and optimality criteria. The same is true
for EPC codes, of which, as we have seen above, LRC codes are a
special case. In particular, one goal is to keep the size of the required
finite field small, since operations over a small field have
less complexity than ones over a larger field due to the smaller look-up
tables involved. For example, Integrated Interleaved
(II) codes~\cite{hapkt,tk}
over $GF(q)$, where $q> \max\{m,n\}$, were proposed in~\cite{bh}
as LRC codes. The construction in~\cite{ll,ll2} (stair codes) reduces field
size when failures are correlated. Extended Integrated Interleaved
(EII) codes~\cite{bh2} unify product codes
and II codes.

As is the case with LRC codes, construction of EPC codes involves
optimality issues. For example,
LRC codes optimizing the minimum distance were presented
in~\cite{tb}. Except for special cases, II codes are
not optimal as LRC codes, but the codes in~\cite{tb} require a field
of size at least $mn$. The same happens with
EII codes:  except for special cases, they do
not optimize the minimum distance~\cite{bh2}.

There are stronger criteria for optimization than the minimum
distance in LRC codes. For example, PMDS
codes~\cite{bhh,bpsy,gybs,ghjy,hy,hsx} satisfy the
Maximally Recoverable (MR) property\cite{ghjy,ghswy}. The
definition of the MR
property is extended for EPC codes in~\cite{ghswy}, but it turns out
that EPC codes with the MR property are difficult to obtain. For
example, in~\cite{ghswy} it was proven that an EPC code
$EP(n,1;n,1;1)$ (i.e., one vertical and one horizontal parity per column
and row and one extra parity) with the MR property requires a field
whose size is superlinear on $n$.

The EII codes presented in~\cite{bh2}, of which II
codes~\cite{bh,tk,w,z2,z3} are special cases, in general assume a
field $GF(q)$ such that $q> \max\{m,n\}$ and the individual codes are
RS type of codes. An exception occurs in~\cite{w}, where binary codes
involving BCH codes are presented. A more general case is given
in~\cite{hys}, where using the techniques of generalized tensor
codes~\cite{if}, the authors extend II codes to any field.
There are also several
publications involving binary codes and codes with different field
sizes with local and global
properties~\cite{flgl,gc,hyus,kno,mg,ns,ska,sz,zy2}.

Although in general EII codes are not optimal with respect to the
minimum distance, they are still
attractive as LRC codes, since they provide versatile alternatives to optimal
codes due to their smaller field size and their iterative row-column decoding
algorithm, allowing them to correct a variety of erasures that are
uncorrectable by traditional LRC codes. An open
problem for EII codes as defined in~\cite{bh2} 
is to extend them
to  codes over any field $GF(q)$, and in particular to
codes over the binary field $GF(2)$. One of the goals of this paper
is to provide such an extension.
The constructions to be presented can be extended to finite fields of any
characteristic but, for simplicity, throughout this paper we assume that
the fields considered have characteristic 2, so, a field $GF(q)$
means $q\eq 2^b$ with $b\geq 1$.

The new contributions are:

\begin{enumerate}

\item We present a new definition of $t$-level EII codes as the direct sum of
certain simple array codes whose rows are codewords in a code over any
field $GF(q)$, in particular, over the binary field $GF(2)$. We call
such component codes 1-level EII codes and we
show that product codes are a special case of them.

\item We give the dimension and a lower bound on the minimum distance
of the new codes. We show that when the component 1-level EII codes
are product codes, in which
case we call the new codes $t$-level EII-PC codes, then the minimum
distance is exactly equal to this lower bound.

\item We show that in many cases, the new codes give better minimum
distance than other LRC codes with the same parameters and we also
show cases in which the upper bound~(\ref{boundcmd}) on the minimum
distance is met. We also study a different parameter to measure the
performance of an LRC code: the average number of erasures that the
code can tolerate, assuming that erasures occur sequentially. We demonstrate
that this parameter and the minimum distance of the code are not
necessarily correlated.

\item We state and prove the error-erasure correcting capability of $t$-level
EII codes and we give recursive systematic encoding and decoding
algorithms for combinations of errors and erasures. Although the
decoding of erasures is unambiguous, we study the problem of
miscorrection when errors and erasures occur. We show that if in some
level of the decoding a miscorrection occurs and one tries to guess
where the miscorrection occurred, then there may be more
than one possible solution to the decoding algorithm.

\item We show that for a $t$-level EII-PC code, the code of
transpose arrays is also a $t$-level EII-PC code.

\item We show that the special case of $t$-level II and EII-PC codes in
which the component codes are RS codes, which we call $t$-level
II and EII-RS codes, corresponds to the cases of
II and EII codes usually studied in literature, showing that our
construction in fact generalizes previous constructions.

\item We give sufficient conditions for a $t$-level EII-PC code allowing
for a systematic encoding with an uniform distribution of
the parity symbols. We show that these
sufficient conditions, in particular, cover the known solution to the problem
for the special case of $t$-level EII-RS codes.

\item We study the
problem of ordering the symbols
in a $t$-level EII-PC code that optimizes the length of a correctable burst.
We improve previously optimal constructions by utilizing both the
code and the code of transpose arrays to order the symbols.

\end{enumerate}

The paper is structured as follows: Section~\ref{sub1level}
gives the definition of 1-level EII codes together with properties and examples,
while Section~\ref{subgen} extends this definition to $t$-level EII codes
over any field $GF(q)$ as a direct sum of $t$ 1-level EII codes. We
also define and give examples
of extended integrated interleaved product codes (EII-PC) as a
special case of EII codes, as well as
the basic properties of EII codes,
like their dimension and a lower bound on their minimum distance.
Section~\ref{suboptimal} presents some special cases of EII codes
meeting bound~(\ref{boundcmd}) and comparisons with known codes.
Section~\ref{encoding} presents a
systematic encoding algorithm and an error-erasure decoding algorithm
that assumes no miscorrections occur. The algorithm is unambiguous for
erasure decoding, but if miscorrections occur when correcting errors
and erasures and the decoding algorithm is adapted by guessing where the
miscorrections occurred, we show that the solution may not be unique.
Section~\ref{subpropEIIPC} concentrates on properties of
EII-PC codes. In particular,
it is shown that the set of transpose arrays of a
$t$-level EII-PC code is also a $t$-level EII-PC code. This property
allows for iterative decoding on rows as well as on columns,
extending the iterative decoding of product codes.
Next we give two applications.
In Section~\ref{Uniform} we consider the problem of uniform
distribution of the parity symbols in an EII-PC code, which was also
treated in~\cite{bh2} for EII-RS codes and in~\cite{chpb} for a family of codes
similar to II codes. In particular, we give a sufficient condition
allowing such uniform distribution.
In Section~\ref{burst},
we study the problem of
reading the symbols in an EII-PC code in an order that maximizes the
length of the longest burst that the code can correct, a problem also
considered in literature~\cite{bcl,mv}, where optimal constructions
were given. However, the horizontal-vertical decoding allows to often
improve this maximal length, a fact that we illustrate with
examples. We end the paper by drawing some conclusions in
Section~\ref{conclusions}.

\section{Definition and Properties of 1-level EII Codes}
\label{sub1level}
Since the components of a $t$-level EII code to be defined in the
next section are 1-level EII codes, we start by defining such codes.
Essentially, the rows of an $m\times n$ array in a 1-level EII code
over $GF(q)$ are in an $[n,n-u_0,d^{\rm (H)}_0]$ code in $GF(q)$, while the
first $n-u_0$ columns are in a vertical $[m,s_0,d^{\rm (V)}_0]$ code
over $(GF(q))^{n-u_0}$. In addition, we may pad the array with a
certain number of 0-columns, a requirement that will be justified in
the next section. Explicitly,

\begin{definition}
\label{defEII1}
{\rm
Take two integers $0\leq u_0<u_1\leq n$ and let
$\uu$ be the following (non-decreasing) vector of length
$m=s_0+s_1$, where $s_0\geq 1$, $s_1\geq 0$ and, if $s_1\eq 0$, $u_1\eq n$:

\begin{eqnarray}
\label{equu1}
\uu &=&
\left(\overbrace{u_0,u_0,\ldots,u_0}^{s_0},\overbrace{u_1,u_1,\ldots,u_1}^{s_1}\right).
\end{eqnarray}

Consider an $[n,n-u_0,d^{\rm (H)}_0]$ code $\C_0$ over a finite field $GF(q)$
that admits a systematic encoder on its first $n-u_0$ symbols and
a (vertical) $[m,s_0,d^{\rm (V)}_0]$ code $\cV_0$ over
$\left(GF(q)\right)^{u_{1}-u_{0}}$ that is linear over $GF(q)$,

We say that an $m\times n$ array $C\eq (c_{i,j})_{\substack{0\leq
i\leq m-1\\ 0\leq j\leq n-1}}$ is in a 1-level EII code $\C(n,\uu)$,
where $\uu$ given by~(\ref{equu1}), if:

\begin{enumerate}

\item $c_{i,j}\eq 0$ for $0\leq i\leq m-1$ and $0\leq j\leq n-u_1-1$.

\item For each $i$, $0\leq i\leq m-1$, vectors $(c_{i,n-u_1},c_{i,n-u_1+1},\ldots,c_{i,n-u_0-1})$ are
the symbols of a codeword in the (vertical) $[m,s_0,d_0^{\rm (V)}]$
code $\cV_0$ over $(GF(q))^{u_1-u_0}$.

\item For each $i$, $0\leq i\leq m-1$,
vector $(c_{i,0},c_{i,1},\ldots,c_{i,n-1})$
is a codeword in $\C_0$.

\end{enumerate}
\qed
}
\end{definition}

Definition~\ref{defEII1} implicitly gives a systematic encoding
algorithm, as illustrated in the next example.

\begin{example}
\label{ex0}
{\em
Consider a $\C(15,(4,4,8,8,8))$ 1-level binary EII code where the horizontal code
$\C_0$ is the $[15,11,3]$ cyclic Hamming code over $GF(2)$ with
generator polynomial $1+x+x^4$
and the vertical code $\cV_0$ is a $[5,2,4]$ (shortened) Reed-Solomon
(RS) code~\cite{ms} over $GF(16)$ with generator polynomial $(x+1)(x+\al)(x+\al^2)$,
where $\al$ is a primitive element in $GF(16)$ such that $1\xor\al\xor\al^4\eq 0$.
Represent a 4-bit symbol $a_0+a_1\al+a_2\al^2+a_3\al^3$ in $GF(16)$ by $(a_0,a_1,a_2,a_3)$.
Below is a systematically encoded $5\times 15$ array, where the data
corresponds to entries $(i,j)$ with $0\leq i\leq 1$ and $7\leq j\leq 10$:

\begin{center}
\begin{tabular}{c|c|c|c|c|c|c|c||c|c|c|c||c|c|c|c|}
\multicolumn{1}{c}{\phantom{0}}&\multicolumn{1}{c}{\scriptsize
0}&\multicolumn{1}{c}{\scriptsize 1}&\multicolumn{1}{c}{\scriptsize 2}&
\multicolumn{1}{c}{\scriptsize 3}& \multicolumn{1}{c}{\scriptsize
4}&\multicolumn{1}{c}{\scriptsize 5}&\multicolumn{1}{c}{\scriptsize
6}&\multicolumn{1}{c}{\scriptsize 7}&\multicolumn{1}{c} {\scriptsize 8}&\multicolumn{1}{c}{\scriptsize 9}&\multicolumn{1}{c}{\scriptsize
10}&\multicolumn{1}{c}{\scriptsize 11}&\multicolumn{1}{c}{\scriptsize 12}&\multicolumn{1}{c}{\scriptsize 13}&\multicolumn{1}{c}{\scriptsize 14}\\
\cline{2-16}
{\scriptsize 0}&0&0&0&0&0&0&0&0&0&0&1&1&0&0&1\\\cline{2-16}
{\scriptsize 1}&0&0&0&0&0&0&0&1&1&0&1&0&0&0&1\\
\cline{9-12}\cline{2-16}
{\scriptsize 2}&0&0&0&0&0&0&0&1&0&1&0&1&1&0&0\\\cline{2-16}
{\scriptsize 3}&0&0&0&0&0&0&0&0&0&0&0&0&0&0&0\\\cline{2-16}
{\scriptsize 4}&0&0&0&0&0&0&0&0&1&1&0&0&1&0&0\\\cline{2-16}
\end{tabular}
\end{center}

First the two symbols $(0,0,0,1)$ and $(1,1,0,1)$, which correspond
to $\al^3$ and $\al^7$ respectively in $GF(16)$, are encoded
systematically into the vertical
$[5,2,4]$ code. Then, each of the 5 symbols, preceded by 7 zeros, are
encoded systematically into the Hamming code, giving the array above.
}
\end{example}

The requirement that the first $n-u_1$ columns of an array in a
1-level EII code are zero seems artificial at this point. If
such a code is taken in isolation, we make $n\eq u_1$, and there are
no zero columns. However, in the next section we will define a
$t$-level EII code as the sum of $t$ 1-level EII codes. We will see
that the zero columns make this sum a direct sum.

The next lemma gives the dimension and a lower bound on the minimum
distance of 1-level EII codes.

\begin{lemma}
\label{l1}
{\em
The 1-level EII code $\C(n\,,\,\uu)$
over $GF(q)$ given by Definition~\ref{defEII1} has
dimension $s_0(u_1-u_0)$ and minimum distance $d\geq d_0^{\rm (V)}d_0^{\rm (H)}$.
}
\end{lemma}

\noindent\pf
Without loss of generality, assume that the data is encoded
systematically. 
The dimension is immediate, since the encoding of the $s_0\times
(u_1-u_0)$ data array into the $m\times n$ array in
$\C(n\,,\,\uu)$
is 1-1.

Now, assume that $d_0^{\rm (V)}d_0^{\rm (H)}-1$ erasures have
occurred. Let $\ell$ be the number of rows with at least $d_0^{\rm
(H)}$ erasures each. Then, $\ell\leq d_0^{\rm (V)}-1$. In each of the
$n-\ell$ remaining rows, there are at most $d_0^{\rm (H)}-1$ erasures. Such
erasures are corrected by the horizontal code $\C_0$. 
Then, the first $n-u_0$ entries of the 
$\ell$ rows with at least $d_0^{\rm (H)}$ erasures each can be corrected by the
vertical code $\cV_0$. Encoding each of these 
corrected 
$\ell$ entries
using the horizontal code $\C_0$, all the erasures are corrected,
hence the result follows.
\qed

The inequality $d\geq d_0^{\rm (V)}d_0^{\rm
(H)}$ from Lemma~\ref{l1} may not become equality in
general. In effect, consider the 1-level binary EII code
$\C(15,(4,4,8,8,8))$ of Example~\ref{ex0}. According to
Lemma~\ref{l1}, the code has dimension $(2)(4)\eq 8$ and
minimum distance $d\geq (4)(3)\eq 12$. If we encode (systematically)
the 255 possible non-zero data arrays, we find out that the
minimum distance of the code is 14. The array depicted in
Example~\ref{ex0} has in fact weight 14.

\begin{example}
\label{ex1}
{\em
Consider a $\C(8,(4,4,8,8,8))$ 1-level binary EII code where the horizontal code
$\C_0$ is an $[8,4,4]$ extended Hamming code over $GF(2)$ and the
vertical code $\cV_0$ is the $[5,2,4]$ shortened RS code over $GF(16)$ of
Example~\ref{ex0}. According to Lemma~\ref{l1}, the code has
dimension 8 and minimum distance $d\geq (4)(4)\eq 16$.

Notice that, in particular,
$\C(8,(4,4,8,8,8))$ is a $(5,8,8;4,4,d,2)$ LRC code by
Definition~\ref{defLRC}, where $d\geq 16$. Consider the
bound~(\ref{boundcmd}) on the minimum distance of a $(5,8,8;4,4,d,2)$ LRC
code. In this case, we have $k^*\eq k_{\rm opt}^{(2)}\,[8,4]\eq 4$,
so $\lceil k/k^*\rceil\eq 2$. In particular, taking $j\eq 1$
in~(\ref{boundcmd}), we obtain $d\leq d_{\rm opt}^{(2)}\,[32\,,\,4]$.

By the Griesmer bound~\cite{gri}, if $d_{\rm
opt}^{(2)}\,[N\,,\,4]\,>\,16$, then $N\geq \sum_{i=0}^3\lceil
17/2^i\rceil\eq 34$. So, $d\leq d_{\rm
opt}^{(2)}\,[32\,,\,4]\leq 16$, showing that $\C(8,(4,4,8,8,8))$
meets bound~(\ref{boundcmd}).
We will generalize this example in Section~\ref{suboptimal}.
}
\end{example}

A special case of a 1-level EII code
$\C(n\,,\,\uu)$ is a product code: in this case, both rows and
columns are individually encoded sharing a parity-check matrix.
Specifically,

\begin{definition}
\label{defEII1PC}
{\rm
Let $\uu$ be given
by~(\ref{equu1}) and let
$h_{x,y}$, where ${0\leq x,y\leq \max\{m,n\}-1}$, be
a set of coefficients in a finite field $GF(q)$,
such that the following are matrices of
rank $s$ for $\max\{v+s,w\}\leq \max\{m,n\}$:

\begin{eqnarray}
\label{Hvws}
H_{s\,,\,w\,,\,v}&=&\left(
\begin{array}{ccccc}
h_{v\,,\,0}&h_{v\,,\,1}&h_{v\,,\,2}&\ldots &h_{v\,,\,w-1}\\
h_{v+1\,,\,0}&h_{v+1\,,\,1}&h_{v+1\,,\,2}&\ldots &h_{v+1\,,\,w-1}\\
\vdots&\vdots&\vdots&\ddots&\vdots\\
h_{v+s-1\,,\,0}&h_{v+s-1\,,\,1}&h_{v+s-1\,,\,2}&\ldots &h_{v+s-1\,,\,w-1}\\
\end{array}
\right).
\end{eqnarray}

Assume that $\C_0$ is an $[n,n-u_0,d_0]$ code whose parity-check
matrix is $H_{u_0\,,\,n\,,\,0}$,
$\cV_0^{(q)}$ is an $[m,s_0,d_0^{\rm (V)}]$ (vertical)
code over $GF(q)$, whose parity-check
matrix is $H_{s_1\,,\,m\,,\,0}$, where $H_{u_0\,,\,n\,,\,0}$
and $H_{s_1\,,\,m\,,\,0}$ are given by~(\ref{Hvws}). Consider the
1-level code $\C(n,\uu)$ with horizontal code $\C_0$
and vertical code $\cV_0$ over
$(GF(q))^{u_1-u_0}$ equal to $(\cV_0^{(q)})^{u_1-u_0}$ according to
Definition~\ref{defEII1}. Then we say that $\C(n,\uu)$ is a 1-level
EII-PC code. If $\C_0$ and $\cV_0^{(q)}$ are MDS codes, we say that
$\C(n,\uu)$ is a 1-level EII-MDS code, while if they are (shortened)
RS codes, we say that it is a 1-level EII-RS code. \qed
}
\end{definition}

If $n\eq u_1$, the 1-level code $\C(n,\uu)$ in Definition~\ref{defEII1PC}, in
particular, is the product code~\cite{ms}
$\C_0\times\cV_0^{(q)}$.
Given an array in a 1-level EII-PC code $\C(n,\uu)$ as defined above,
it is well
known that each of the last $u_0$ columns is also in $\cV_0^{(q)}$
and the minimum distance of
$\C(n,\uu)$ is $d\eq d_0^{\rm (V)}d_0^{\rm (H)}$~\cite{ms}.

\begin{example}
\label{ex2}
{\em Consider the binary 1-level EII-PC
code $\C(7,(3,3,3,3,7,7,7))$
with $\C_0\eq\cV_0^{(2)}$ the $[7,4,3]$ Hamming code over
$GF(2)$ as given by Definition~\ref{defEII1PC}.
Then, $\C(7,(3,3,3,3,7,7,7))$ is the product code
$\C_0\times\C_0$, which has
minimum distance $d\eq (3)(3)\eq 9$. Let us denote this code
by $\C^{(b)}$. We want to compare it next with a 1-level EII
code $\C(7,(3,3,3,3,7,7,7))$ that is not a product code.

In effect, let $\C^{(a)}$ be a binary 1-level EII code $\C(7,(3,3,3,3,7,7,7))$ with
$\C_0$ the $[7,4,3]$ Hamming code over $GF(2)$ and
$\cV_0$ the $[7,4,4]$ (shortened) RS code over $GF(16)$ as given by
Definition~\ref{defEII1}.
As in Example~\ref{ex1}, its minimum distance is $d\geq
(4)(3)\eq 12$.

Notice that $\C^{(a)}$ is a $[49,16,d]$ code with $d\geq 12$, while
$\C^{(b)}$ is a $[49,16,9]$ code. So, $\C^{(a)}$ has larger minimum
distance. However, there are
patterns that can be corrected by $\C^{(b)}$ and not by $\C^{(a)}$.
For example, take the following array, where E
denotes erasure:

\begin{center}
\begin{tabular}{|c|c|c|c||c|c|c|}
\hline
\phantom{E}&E&E&\phantom{E}&\phantom{E}&E&E\\
\hline
E&\phantom{E}&E&\phantom{E}&\phantom{E}&E&E\\
\hline
\phantom{E}&\phantom{E}&\phantom{E}&\phantom{E}&\phantom{E}&\phantom{E}&\phantom{E}\\
\hline
\phantom{E}&\phantom{E}&\phantom{E}&\phantom{E}&\phantom{E}&\phantom{E}&\phantom{E}\\
\hline\hline
E&E&E&E&E&E&\phantom{E}\\
\hline
\phantom{E}&\phantom{E}&\phantom{E}&\phantom{E}&\phantom{E}&\phantom{E}&\phantom{E}\\
\hline
\phantom{E}&\phantom{E}&E&E&E&E&\phantom{E}\\
\hline
\end{tabular}
\end{center}

This pattern is uncorrectable by $\C^{(a)}$. In
effect, none of the codewords can be corrected by the horizontal
code, since they have more than two erasures each. The vertical code over
$GF(16)$ cannot correct the four symbols having erasures either, since
this code can correct at most three erasures. However $\C^{(b)}$ can
correct all the columns with up to two erasures, leaving four rows
with two erasures each, which are correctable.

Let us take another measure for the performance of a code, which was
already considered in~\cite{bh}.
Assume that, given an erasure-correcting code, erasures occur one
after the other until the code
encounters an uncorrectable erasure pattern. How many erasures can
the code correct on average? This parameter is intimately related to
the mean time to data loss (MTTDL) of the system (the relevance of
MTTDL was also stated in~\cite{ska}). The exact
solution is related to
``birthday surprise'' type of problems~\cite{gm,gm2,kn}, but a Monte
Carlo simulation shows that on average,
$\C^{(a)}$ can correct 17.8 erasures, while $\C^{(b)}$ can correct
22.7 erasures, in spite that the minimum distance of $\C^{(b)}$ is
smaller than the minimum distance of $\C^{(a)}$. In the simulation,
we are not taking into account that the minimum distance of
$\C^{(a)}$ may be larger than 12. In order to enhance the iterative
decoding algorithm, uncorrectable patterns by this algorithm may be
checked using the parity-check matrices of the codes. In any case,
if the main parameter to evaluate the performance of an
erasure-correcting code is the average number of erasures until an
uncorrectable pattern is encountered, then the 1-level EII-PC code
$\C^{(b)}$ looks attractive with respect to the 1-level EII
code $\C^{(a)}$.
}
\end{example}

Consider a 1-level EII-PC code
$\C(n\,,\,(\overbrace{u_0,u_0,\ldots,u_0}^{s_0},\overbrace{n,n,\ldots,n}^{m-s_0}))$.
The rows of a transpose array in the code are the columns of the
array, so the set of transpose arrays constitute a 1-level EII-PC code
$\C(m\,,\,(\overbrace{s_0,s_0,\ldots,s_0}^{n-u_0},\overbrace{m,m\ldots,m}^{u_0}))$.
This observation is well known, but we will generalize it later for $t$-level
EII-PC codes, so we state it as a lemma.

\begin{lemma}
\label{l2}
{\em
Consider a 1-level EII-PC code
$\C(n\,,\,(\overbrace{u_0,u_0,\ldots,u_0}^{s_0},\overbrace{n,n,\ldots,n}^{m-s_0}))$
over $GF(q)$ as given by Definition~\ref{defEII1PC}.
Then, the set of transpose arrays of the
arrays in the code is a 1-level EII-PC code
$\C(m\,,\,(\overbrace{s_0,s_0,\ldots,s_0}^{n-u_0},\overbrace{m,m\ldots,m}^{u_0}))$.
\qed
}
\end{lemma}

\section{Definition and Properties of $t$-level EII Codes}
\label{subgen}
Now we are ready to give a general definition of $t$-level EII codes.
Essentially, a $t$-level EII code is a direct sum of $t$ $1$-level
EII codes. Explicitly:

\begin{definition}
\label{defEII}
{\rm
Given $t>1$, take $t+1$ integers $0\leq u_0<u_1<\ldots <u_{t-1}<u_t\leq n$ and let
$\uu$ be the following (non-decreasing) vector of length
$m=s_0+s_1+\cdots +s_{t-1}+s_t$, where $s_i\geq 1$ for $0\leq
i\leq t-1$, $s_t\geq 0$ and, if $s_t\eq 0$, $u_t\eq n$:

\vspace{-3mm}

\begin{eqnarray}
\label{equu}
\uu &=&
\left(\overbrace{u_0,u_0,\ldots,u_0}^{s_0},\overbrace{u_1,u_1,\ldots,u_1}^{s_1},\ldots,
\overbrace{u_{t-1},u_{t-1},\ldots,u_{t-1}}^{s_{t-1}},\overbrace{u_t,u_t,\ldots,u_t}^{s_{t}}\right).
\end{eqnarray}

Let

\vspace{-3mm}

\begin{eqnarray}
\label{hsi}
\hs_i &=&
\sum_{j=i}^{t}s_j\quad {\rm for}\quad 0\leq i\leq t.
\end{eqnarray}

\vspace{-3mm}

Consider a set $\{\C_i\}$, $0\leq i\leq t-1$, of $t$ linear
$[n,n-u_i,d^{\rm (H)}_i]$ nested codes over a finite field $GF(q)$
$\C_{t-1}\subset \C_{t-2}\subset \ldots
\subset\C_0$, where each $\C_i$ admits a systematic encoder on its
first $n-u_i$ symbols,
and a set $\{\cV_i\}$, $0\leq i\leq t-1$, of $t$ (vertical)
codes, where each $\cV_{i}$ is an
$[m,m-\hs_{t-i},d^{\rm (V)}_i]$ code over
$\left(GF(q)\right)^{u_{t-i}-u_{t-i-1}}$ that is linear over $GF(q)$.
For each $i$ such that $0\leq i\leq t-1$, consider the 1-level EII
code $\C(n,\uu^{(i)})$ as given by Definition~\ref{defEII1}, where

\vspace{-3mm}

\begin{eqnarray}
\label{equui}
\uu^{(i)}\eq
(\overbrace{u_i,u_i,\ldots,u_i}^{m-\hs_{i+1}},\overbrace{u_{i+1},u_{i+1},\ldots,u_{i+1}}^{\hs_{i+1}}),
\end{eqnarray}
the horizontal code is $\C_{i}$ and the
vertical code is $\cV_{t-1-i}$. Let

\vspace{-3mm}

\begin{eqnarray}
\label{eqCnu}
\C(n,\uu)&\eq &\bigoplus_{i=0}^{t-1}\,\C(n,\uu^{(i)}).
\end{eqnarray}

\vspace{-2mm}

Then we say that $\C(n,\uu)$ is a $t$-level EII code. If $s_t\eq 0$, we say that $\C(n,\uu)$ is a
$t$-level II code.
\qed
}
\end{definition}

\begin{example}
\label{ex3}
{\em
Consider 2-level EII codes. According to
Definition~\ref{defEII}, we take 3 integers $0\leq u_0<u_1<u_2\leq
n$, and
$$\uu\eq\left(\overbrace{u_0,u_0,\ldots,u_0}^{s_0},\overbrace{u_1,u_1,\ldots,u_1}^{s_1},\overbrace{u_2,u_2,\ldots,u_2}^{s_{2}}\right),$$
where $s_0,s_1\geq 1$, $s_2\geq 0$, $u_2\eq n$ if $s_2\eq 0$ and $s_0+s_1+s_2\eq m$.
Consider the linear (horizontal) codes $\C_{1}\subset\C_0$,
where $\C_1$ is an $[n,n-u_1,d^{\rm (H)}_1]$ code over $GF(q)$ and $\C_0$ is an
$[n,n-u_0,d^{\rm (H)}_0]$ code over $G(q)$, and the (vertical) codes
$\cV_0$ and $\cV_1$, where $\cV_0$  is an $[m,s_0+s_1,d^{\rm (V)}_0]$ code over
$\left(GF(q)\right)^{u_{2}-u_1}$ and $\cV_1$ is an $[m,s_0,d^{\rm (V)}_1]$ code over
$\left(GF(q)\right)^{u_{1}-u_0}$.

According to Definition~\ref{defEII}, if $C$ is an array in
$\C(n,\uu)$, then $C\eq C_1\xor C_0$, where
$C_1\in\C(n,(\overbrace{u_1,u_1,\ldots,u_1}^{s_0+s_1},\overbrace{u_2,u_2,\ldots,u_2}^{s_{2}}))$
with horizontal code $\C_1$ and vertical code $\cV_0$, and
$C_0\in\C(n,(\overbrace{u_0,u_0,\ldots,u_0}^{s_0},\overbrace{u_1,u_1,\ldots,u_1}^{s_1+s_2}))$
with horizontal code $\C_0$ and vertical code $\cV_1$.

For example, consider a $\C(15,(4,4,8,15))$ 2-level EII code over the binary field
$GF(2)$ (which consists of $4\times 15$ arrays), such that
$\C_{1}\subset\C_0$, where
$\C_0$ is a $[15,11,3]$ Hamming code and $\C_1$ is a $[15,7,5]$ BCH
code, while $\cV_0$ is a $[4,3,2]$ vertical code over $(GF(2))^{7}$
(for example, by taking an even parity code) and
$\cV_1$ is a $[4,2,3]$ vertical code over $(GF(2))^{4}$ (for example,
by taking a $[4,2,3]$ shortened RS code over $GF(16)$).

If $C\in\C(15,(4,4,8,15))$, then $C\eq C_1\xor C_0$, where
$C_1\in\C(15,(8,8,8,15))$ and
$C_0\in\C(15,(4,4,8,8))$.
So, denoting data by $D$ and parity by $P$,
we have

\begin{eqnarray*}
C_1&=&
\begin{array}{|c|c|c|c|c|c|c||c|c|c|c|c|c|c|c|}
\hline
D&D&D&D&D&D&D&P&P&P&P&P&P&P&P\\
\hline
D&D&D&D&D&D&D&P&P&P&P&P&P&P&P\\
\hline
D&D&D&D&D&D&D&P&P&P&P&P&P&P&P\\
\hline\hline
P&P&P&P&P&P&P&P&P&P&P&P&P&P&P\\
\hline
\end{array}
\end{eqnarray*}

\begin{eqnarray*}
C_0&=&
\begin{array}{|c|c|c|c|c|c|c||c|c|c|c||c|c|c|c|}
\hline
\hspace{.48mm}0\hspace{.48mm}&\hspace{.48mm}0\hspace{.48mm}&\hspace{.48mm}0\hspace{.48mm}&\hspace{.48mm}0\hspace{.48mm}&
\hspace{.48mm}0\hspace{.48mm}&\hspace{.48mm}0\hspace{.48mm}&\hspace{.48mm}0\hspace{.48mm}&D&D&D&D&P&P&P&P\\
\hline
0&0&0&0&0&0&0&D&D&D&D&P&P&P&P\\
\hline\hline
0&0&0&0&0&0&0&P&P&P&P&P&P&P&P\\
\hline
0&0&0&0&0&0&0&P&P&P&P&P&P&P&P\\
\hline
\end{array}
\end{eqnarray*}
and $C\eq C_1\xor C_0$. This example also shows a natural way of
encoding the data, although this encoding is not systematic.
}
\end{example}

\begin{example}
\label{ex4}
{\em
Consider 3-level EII codes. According to
Definition~\ref{defEII},
we take 3 integers $0\leq u_0<u_1<u_2<u_3\leq n$, and let
$$\uu\eq\left(\overbrace{u_0,u_0,\ldots,u_0}^{s_0},\overbrace{u_1,u_1,\ldots,u_1}^{s_1}
,\overbrace{u_2,u_2,\ldots,u_2}^{s_{2}},\overbrace{u_3,u_3,\ldots,u_3}^{s_{3}}\right)$$
where
$s_0,s_1,s_2\geq 1$, $s_3\geq 0$, $u_3\eq n$ if $s_3\eq 0$ and $s_0+s_1+s_2+s_3\eq m$.
Consider the linear codes $\C_{2}\subset\C_{1}\subset\C_0$,
where $\C_2$ is an $[n,n-u_2,d^{\rm (H)}_2]$ code,
$\C_1$ is an $[n,n-u_1,d^{\rm (H)}_1]$ code and $\C_0$ is an
$[n,n-u_0,d^{\rm (H)}_0]$ code over $G(q)$, while the (vertical) code
$\cV_0$ is an $[m,s_0+s_1+s_2,d^{\rm (V)}_0]$ code over
$\left(GF(q)\right)^{u_{3}-u_2}$,
$\cV_1$ is an $[m,s_0+s_1,d^{\rm (V)}_1]$ code over
$\left(GF(q)\right)^{u_{2}-u_1}$, and $\cV_2$ is an $[m,s_0,d^{\rm (V)}_2]$ code over
$\left(GF(q)\right)^{u_{1}-u_0}$.

Then, according to Definition~\ref{defEII},
if $C$ is an array in
$\C(n,\uu)$, then $C\eq C_2\xor C_1\xor C_0$, where\\
$C_2\in
\C(n,(\overbrace{u_2,u_2,\ldots,u_2}^{s_0+s_1+s_2},\overbrace{u_3,u_3,\ldots,u_3}^{s_{3}}))$
with horizontal code $\C_2$ and vertical code $\cV_0$,\\
$C_1\in\C(n,(\overbrace{u_1,u_1,\ldots,u_1}^{s_0+s_1},\overbrace{u_2,u_2,\ldots,u_2}^{s_{2}+s_3}))$
with horizontal code $\C_1$ and vertical code $\cV_1$ and\\
$C_0\in\C(n,(\overbrace{u_0,u_0,\ldots,u_0}^{s_0},\overbrace{u_1,u_1,\ldots,u_1}^{s_1+s_2+s_3}))$
with horizontal code $\C_0$ and vertical code $\cV_2$. 

For example, take
a 3-level binary EII code $\C(15,(4,4,8,10,15))$ such that
$\C_2\subset\C_{1}\subset\C_0$, where
$\C_0$ is a $[15,11,3]$ Hamming code, $\C_1$ is a $[15,7,5]$ BCH
code and $\C_2$ is a $[15,5,7]$ BCH
code, while $\cV_0$ is a $[5,4,2]$ parity code over $(GF(2))^{5}$,
$\cV_1$ is a $[5,3,3]$ doubly extended RS code over $GF(4)$ and
$\cV_2$ is a $[5,2,4]$ shortened RS code over $GF(16)$.

If $C\in\C(15,(4,4,8,10,15))$, then $C\eq C_2\xor C_1\xor C_0$, where
$C_2\in\C(15,(10,10,10,10,15))$,
$C_1\in\C(15,(8,8,8,10,10))$ and
$C_0\in\C(15,(4,4,8,8,8))$.
Proceeding as in Example~\ref{ex3}, we obtain a non-systematic encoder.
}
\end{example}

If $u_0\eq 0$ in Definition~\ref{defEII}, then the code $\C_0$ is an
$[n,n,1]$ code, that is, the whole space, with no erasure-correcting
capabilities. Similarly, if $s_t\eq 0$ (that is, an II code), then
$\cV_0$ is an $[m,m,1]$ code over $(GF(q))^{n-u_{t-1}}$, also the
whole space.
Also notice that if $C$ is an $m\times n$ array in a code $\C(n,\uu)$
as given by Definition~\ref{defEII}, since $\C_i\subset\C_0$ for
$i>0$, then each row in $C$ is in $\C_0$.

Next we prove that the sum of 1-level EII codes given
by~(\ref{eqCnu}) is in fact a direct sum, justifying the assumption
of padding with zeros the first $n-u_1$ columns in
Definition~\ref{defEII1}.

\begin{lemma}
\label{cor0}
{\em
Consider the $t$-level EII code $\C(n,\uu)$ given by
Definition~\ref{defEII} and assume that $t>1$. Then
$\C(n,\uu)$ is the direct sum of
$\C(n,\uu^{(0)})$ and $\C(n,\uu')$, where
\begin{eqnarray}
\label{equu'}
\uu' &=&
\left(\overbrace{u_1,u_1,\ldots,u_1}^{s_0+s_1},\overbrace{u_2,u_2,\ldots,u_2}^{s_2},\ldots,
\overbrace{u_{t-1},u_{t-1},\ldots,u_{t-1}}^{s_{t-1}},\overbrace{u_t,u_t,\ldots,u_t}^{s_{t}}\right).
\end{eqnarray}
}
\end{lemma}

\vspace{-3mm}

\noindent\pf
By~(\ref{eqCnu}),
$\C(n,\uu')\eq\bigoplus_{i=1}^{t-1}\,\C(n,\uu^{(i)})$, so, $\C(n,\uu)$ is the sum of
$\C(n,\uu^{(0)})$ and $\C(n,\uu')$. In order to prove that it is also
a direct sum, we have to
show that the intersection between $\C(n,\uu^{(0)})$ and
$\C(n,\uu')$ is the zero array.

In effect, take $C\in\C(n,u^{(0)})\cap\C(n,\uu')$.
Let $C\eq (c_{i,j})_{\substack{0\leq
i\leq m-1\\ 0\leq j\leq n-1}}$. Since $C\in\C(n,u^{(0)})$,
by Definition~\ref{defEII}, $c_{i,j}\eq 0$ for $0\leq j\leq n-u_{1}-1$.
Since $C\in\C(n,\uu')$, in particular, each row $\uc_i\eq
(c_{i,0},c_{i,1},\ldots c_{i,n-1})$ is in $\C_1$, which is an $[n,n-u_1]$
code. Since the first $n-u_1$ entries of $\uc_i$ are zero, encoding
systematically such first $n-u_1$ entries, we obtain that
$\uc_i$ is zero, so $C$ is the zero array. \qed

\vspace{-3mm}

As done in Section~\ref{sub1level} with 1-level EII codes, we may
assume that the
component 1-level EII codes in Definition~\ref{defEII} are product
codes. We next extend Definition~\ref{defEII1PC} of 1-level EII-PC
codes to $t$-level EII-PC codes.

\begin{definition}
\label{defEIIPC}
{\rm
Consider $t$ and $\uu$ as in Definition~\ref{defEII}.
Take a set of coefficients $h_{x,y}$ in a finite field $GF(q)$
as in Definition~\ref{defEII1PC},
$t$ nested horizontal codes $\C_i$,
$0\leq i\leq t-1$, with parity-check matrix $H_{u_i\,,\,n\,,\,0}$,
$t$ vertical codes $\cV_i$ over $(GF(q))^{u_{t-i}-u_{t-i-1}}$, $0\leq
i\leq t-1$, defined as $\cV_i\eq
(\cV_i^{(q)})^{u_{t-i}-u_{t-i-1}}$, where the codes $\cV_i^{(q)}$
are nested codes over $GF(q)$
with parity-check matrix $H_{\hs_{t-i}\,,\,m\,,\,0}\,$, both
$H_{u_i\,,\,n\,,\,0}$ and $H_{\hs_{t-i}\,,\,m\,,\,0}\,$ given by~(\ref{Hvws}).
Then we say that $\C(n,\uu)$ is a $t$-level EII-PC code. If $s_t\eq 0$, we
say that $\C(n,\uu)$ is a $t$-level II-PC code. If the codes $\C_i$
and $\cV_i^{(q)}$, $0\leq i\leq t-1$, are MDS codes, then we say that
$\C(n,\uu)$ is an EII-MDS code (resp., II-MDS code if $s_t\eq 0$),
while if they are RS codes, we say that $\C(n,\uu)$
is an EII-RS code (resp., II-RS code if $s_t\eq 0$).
\qed
}
\end{definition}

\vspace{-3mm}

\begin{example}
\label{ex20}
{\em
Consider the binary 3-level II-PC code $\C(8,(1,1,1,1,4,4,4,7))$,
where $\C_0$ is the $[8,7,2]$ parity-check code, $\C_1$ is
the $[8,4,4]$ extended Hamming code, $\C_2$ is the $[8,1,8]$
repetition code, $\cV_0^{(2)}$ is the
$[8,8,1]$ code corresponding to the whole space $(GF(2))^8$,
$\cV_1^{(2)}$ is the $[8,7,2]$ parity-check code and
$\cV_2^{(2)}$ is the $[8,4,4]$ extended Hamming code.
We assume that the coefficients $h_{x,y}$ in
Definition~\ref{defEII} are given by the $8\times 8$ matrix

\begin{eqnarray}
\label{eqH}
H_{8,8,0}&=&\left(
\begin{array}{cccccccc}
1&1&1&1&1&1&1&1\\
1&1&0&1&1&0&0&0\\
1&0&1&1&0&1&0&0\\
0&1&1&1&0&0&1&0\\
0&0&0&0&1&0&0&1\\
0&0&0&0&0&1&0&1\\
0&0&0&0&0&0&1&1\\
0&0&0&0&0&0&0&1\\
\end{array}
\right),
\end{eqnarray}
$H_{1,8,0}$ is the parity-check matrix of $\C_0$ and
$\cV_1^{(2)}$, $H_{4,8,0}$ is the parity-check matrix of $\C_1$ and
$\cV_2^{(2)}$ and $H_{7,8,0}$ is the parity-check matrix of $\C_2$, where
$H_{s\,,\,w\,,\,v}$ is given by~(\ref{Hvws}).

According to Definition~\ref{defEIIPC},

\begin{eqnarray*}
\C(8,(1,1,1,1,4,4,4,7))&\eq
&\C(8,(1,1,1,1,4,4,4,4))\xor\C(8,(4,4,4,4,4,4,4,7))\xor\C(8,(7,7,7,7,7,7,7,7)).
\end{eqnarray*}
}
\end{example}

\begin{theorem}
\label{theo1}
{\em
Consider a $t$-level EII code $\C(n,\uu)$ as given by
Definition~\ref{defEII}. Then, $\C(n,\uu)$ has dimension
$mu_t-\sum_{i=0}^{t}s_iu_i$ and minimum distance $d\geq \min\{d_i^{\rm
(H)}d_{t-1-i}^{\rm (V)}\,:\,0\leq i\leq t-1\}$.
}
\end{theorem}

\vspace{-1mm}

\noindent\pf
We proceed by induction on $t$ for the dimension. If $t\eq 1$, by
Lemma~\ref{l2}, the dimension of the code is $s_0(u_1-u_0)$. But
$s_0(u_1-u_0)\eq (s_0+s_1)u_1-s_0u_0-s_1u_1\eq mu_1-s_0u_0-s_1u_1$
and the result follows.

Assume that $t>1$. By Lemma~\ref{cor0},
$\C(n,\uu)$ is the direct sum of
$\C(n,\uu^{(0)})$ and $\C(n,\uu')$, where $\uu'$ is given by~(\ref{equu'}),
so its dimension is the sum of the dimensions of  $\C(n,\uu')$ and of
$\C(n,\uu^{(0)})$.
By induction, since $\C(n,\uu')$ is a $(t-1)$-level EII code, its
dimension is $mu_t-(s_0+s_1)u_1-\sum_{i=2}^ts_iu_i$, while the
dimension of
$\C(n,\uu^{(0)})$,
by Lemma~\ref{l1}, is $s_0(u_1-u_0)$. Adding these two numbers, we
obtain the dimension of the code, as claimed.

Regarding the lower bound on the minimum distance, it will be proven
as Corollary~\ref{cor6} after we state
the erasure correcting capability of a $t$-level EII code in Theorem~\ref{theo2}.
\qed

\vspace{-3mm}

The next corollary is immediate.

\begin{corollary}
\label{cor5}
{\em
Consider a $t$-level EII code $\C(n,\uu)$ as given by
Definition~\ref{defEII}, and assume that $u_t\eq n$, the (horizontal)
codes $\C_i$ are MDS
codes over $GF(q)$ and the (vertical) codes $\cV_{t-i-1}$
are MDS codes over $GF(q^{u_{i+1}-u_i})$, where $0\leq i\leq t-1$.
Then, $\C(n,\uu)$ has dimension
$mn-\sum_{i=0}^{t}s_iu_i$ and minimum distance $d\geq
\min\{(u_i+1)(\hs_{i+1}+1)\;\;{\rm for}\;\;0\leq i\leq t-1\}$. \qed
}
\end{corollary}


\vspace{-3mm}

If we assume that the MDS codes in Corollary~\ref{cor5} are doubly
extended (shortened) RS codes, then
$n\leq q+1$. Also, defining\\
$x\eq \min\{u_{i+1}-u_i\}\;\;{\rm for}\;\;0\leq i\leq t-3$ if
$\hs_{t-1}\eq 1$,
$x\eq \min\{u_{i+1}-u_i\}\;\;{\rm for}\;\;0\leq i\leq t-2$ if
$\hs_{t-1}> 1$ and $s_t\leq 1$ and
$x\eq \min\{u_{i+1}-u_i\}\;\;{\rm for}\;\;0\leq i\leq t-1$ if
$s_{t}> 1$, where $u_t\eq n$, then
$m\leq q^x+1$. If $x>1$, then $m$ may be considerably larger than
$n$, an observation also made in Example~4 of~\cite{hys}. We will
compare EII codes with II codes under the conditions of
Corollary~\ref{cor5} in Example~\ref{ex25} of the next section.

\section{EII Codes Achieving Bound~(\ref{boundcmd}) and
Comparison with Other Codes}
\label{suboptimal}

Since $t$-level II codes (i.e., $s_t\eq 0$ in
Definition~\ref{defEII}) meeting bound~(\ref{boundcmd}) were
presented in~\cite{hys,hyus}, in this section we concentrate on EII
codes with $s_t>0$. We start with a
construction of 1-level EII codes that generalizes Example~\ref{ex1}
and give a sufficient condition under which it meets bound~(\ref{boundcmd}).

\begin{lemma}
\label{l6}
{\em
Consider a $1$-level EII code $\C(n,\uu)$ over
$GF(q)$ as given by Definition~\ref{defEII1} with $u_1\eq n$,
$n-u_0\eq k_{\rm opt}^{(q)}\,[n,d_0^{\rm (H)}]$ and $d_0^{\rm (H)}\eq
d_{\rm opt}^{(q)}\,[n,n-u_0]$.
Assume that there is a
$j$, $0\leq j\leq s_0-1$, such that $d_{\rm
opt}^{(q)}\,[(m-j)n,(s_0-j)(n-u_0)]\eq d_0^{\rm (V)}d_0^{\rm (H)}$.
Then the minimum distance $d$ of $\C(n,\uu)$ meets the bound given
by~(\ref{boundcmd}).
}
\end{lemma}

\noindent\pf
By Lemma~\ref{l1}, the minimum distance $d$ of $\C(n,\uu)$ satisfies
$d\geq d_0^{\rm (V)}d_0^{\rm (H)}$.
By the choice of the code $\C_0$ in the construction,
$k^*\eq n-u_0$. Since the code has dimension $k\eq s_0(n-u_0)$,
then $\lceil k/k^*\rceil\eq s_0$.

Since there is a
$j$, $0\leq j\leq s_0-1$, such that
$d_{\rm
opt}^{(q)}\,[(m-j)n,(s_0-j)(n-u_0)]\eq d_0^{\rm (V)}d_0^{\rm (H)}$,
from~(\ref{boundcmd}), we obtain $d\leq d_0^{\rm (V)}d_0^{\rm
(H)}$, proving the result.
\qed

\vspace{-3mm}



\begin{example}
\label{ex18}
{\em Consider the 1-level EII code
$\C(16,(\overbrace{9,9,\ldots,9}^{m-1},16))$ over $GF(2)$ such that $\C_0$ is a
$[16,7,6]$ extended BCH code and $\cV_0$ is a parity $[m,m-1,2]$ code
over $(GF(2))^{7}$ (hence $m$ can take any value). According to Lemma~\ref{l6} and
since $d_{\rm opt}^{(2)}\,[32,7]\eq 12\eq 2d_{\rm opt}^{(q)}\,[16,7]$~\cite{gr},
$\C(16,(\overbrace{9,9,\ldots,9}^{m-1},11))$ meets
bound~(\ref{boundcmd}) with minimum distance $d\eq 12$.
}
\end{example}

The next lemma gives a specific instance where the construction of
Lemma~\ref{l6} always meets bound~(\ref{boundcmd}).

\begin{lemma}
\label{l6tris}
{\em
Consider a $1$-level EII code 
$\C(q+1,\uu)$ over $GF(q)$ as given by Definition~\ref{defEII1} such
that $m\leq q^2+1$, $2\leq s_1\leq m-1$, $u_0\eq q-1$, $\C_0$ a $[q+1,2,q]$
doubly extended RS code over
$GF(q)$ and $\cV_0$ an $[m,m-s_1,s_1+1]$ (shortened) doubly extended RS code
over $GF(q^2)$. Then,  $\C(q+1,\uu)$ meets
bound~(\ref{boundcmd}) with minimum distance $d\eq q(s_1+1)$.
}
\end{lemma}

\noindent\pf
By Lemma~\ref{l1}, $\C(q+1,\uu)$ has dimension $k\eq 2(m-s_1)$ and
its  minimum distance $d$ satisfies
$d\geq d_0^{\rm (V)}d_0^{\rm (H)}\eq q(s_1+1)$.
Code $\C_0$ is MDS, thus $k^*\eq 2$ and hence $\lceil k/k^*\rceil\eq m-s_1$.

Taking $j\eq m-s_1-1$, by~(\ref{boundcmd}), the result would be proven if we show that
$d_{\rm opt}^{(q)}\,[(q+1)(s_1+1),2]\leq q(s_1+1)$. Assume that this is not the case, so
$d_{\rm opt}^{(q)}\,[(q+1)(s_1+1),2]\geq q(s_1+1)+1$. By the Griesmer
bound, the length $N$ of an $(N,2)$ code over $GF(q)$ with this minimum distance
satisfies

\begin{eqnarray*}
N&\geq &q(s_1+1)+1\,+\,\left\lceil \frac{q(s_1+1)+1}{q}\right\rceil\;\eq\;(q+1)(s_1+1)+2,
\end{eqnarray*}
which is a contradiction.
\qed

\vspace{-3mm}

\begin{example}
\label{ex30}
{\em Consider the 1-level EII code
$\C(5,(\overbrace{3,3,\ldots,3}^{8},\overbrace{5,5,\ldots,5}^{9}))$ over $GF(4)$ such that $\C_0$ is a
$[5,2,4]$ doubly extended RS over $GF(4)$ and $\cV_0$ is a  $[17,8,10]$
doubly extended RS code over $GF(16)$. This one is then an $[85,16,40]$ code over $GF(4)$
whose minimum distance $d\eq 40$, according to Lemma~\ref{l6tris}, meets bound~(\ref{boundcmd}).
}
\end{example}

\vspace{-3mm}

Table~\ref{t1} gives the parameters of some codes obtained with the
construction of Lemma~\ref{l6}. The code $\C_0$ is an $[n,n-u_0,d_0^{\rm (H)}]$
code over $GF(q)$ while the code $\cV_0$ is an $[m,s_0,d_0^{\rm
(V)}]$ code over $GF(q^{n-u_0})$. The minimum distance of the code is
$d\geq d_0^{\rm (V)}d_0^{\rm (H)}$. We indicate with an asterisk when
$d$ achieves bound~(\ref{boundcmd}).

\begin{table}
\begin{center}
\begin{tabular}{|c|c|c|c|c|c|c|c|}
\hline
$m$&$n$&$q$&$u_0$&$s_0$&$\C_0$&$\cV_0$&$d\geq$\\
\hline\hline
$m$&16&2&9&$m-1$&$[16,7,6]$&$[m,m-1,2]$&$12^*$\\
\hline
$m$&16&2&11&$m-1$&$[16,5,8]$&$[m,m-1,2]$&$16^*$\\
\hline
$\leq 33$&16&2&11&$m-2$&$[16,5,8]$&$[m,m-2,3]$&$24^*$\\
\hline
$\leq 33$&16&2&11&$m-3$&$[16,5,8]$&$[m,m-3,4]$&$32^*$\\
\hline
$\leq 33$&16&2&11&$m-7$&$[16,5,8]$&$[m,m-7,8]$&$64^*$\\
\hline
$m$&32&2&26&$m-1$&$[32,6,16]$&$[m,m-1,2]$&$32^*$\\
\hline
$\leq 65$&32&2&26&$m-9$&$[32,6,16]$&$[m,m-9,10]$&$160^*$\\
\hline
$\leq 9$&4&2&1&$m-2$&$[4,3,2]$&$[m,m-2,3]$&$6^*$\\
\hline
$\leq 9$&4&2&1&$m-3$&$[4,3,2]$&$[m,m-3,4]$&$8^*$\\
\hline
$\leq 5$&3&2&1&$m-2$&$[3,2,2]$&$[m,m-2,3]$&$6^*$\\
\hline
$\leq 5$&3&2&1&$m-3$&$[3,2,2]$&$[m,m-3,4]$&$8^*$\\
\hline
$5$&3&2&1&$3$&$[3,2,2]$&$[5,3,3]$&$6^*$\\
\hline
$21$&3&2&1&$18$&$[3,2,2]$&$[21,18,3]$&$6$\\
\hline
$17$&3&2&1&$13$&$[3,2,2]$&$[17,13,4]$&$8$\\
\hline
$41$&3&2&1&$36$&$[3,2,2]$&$[41,36,4]$&$8$\\
\hline
$126$&3&2&1&$120$&$[3,2,2]$&$[126,120,4]$&$8$\\
\hline
$11$&3&2&1&$6$&$[3,2,2]$&$[11,6,5]$&$10$\\
\hline
$21$&3&2&1&$15$&$[3,2,2]$&$[21,15,5]$&$10$\\
\hline
$43$&3&2&1&$36$&$[3,2,2]$&$[43,36,5]$&$10$\\
\hline
$85$&3&2&1&$77$&$[3,2,2]$&$[85,77,5]$&$10$\\
\hline
$m$&5&4&3&$m-1$&$[5,2,4]$&$[m,m-1,2]$&$8^*$\\
\hline
$17$&5&4&3&$1\leq j\leq 15$&$[5,2,4]$&$[m,j,m-j+1]$&$4(m-j+1)^*$\\
\hline
\end{tabular}
\end{center}
\caption{\label{t1} Parameters of some codes obtained with the
construction of Lemma~\ref{l6}. $\cV_0$ is an $[m,s_0,d_0^{\rm
(V)}]$ code over $GF(q^{n-u_0})$. The asterisk means that
bound~(\ref{boundcmd}) is achieved.}
\end{table}


The next lemma gives some 2-level EII codes meeting bound~(\ref{boundcmd}).

\begin{lemma}
\label{l6bis}
{\em
Consider a $2$-level EII code $\C(n,\uu)$ over
$GF(q)$ as given by Definition~\ref{defEII} such that $s_2\eq 1$,
$n-u_0\eq k_{\rm opt}^{(q)}\,[n,d_0^{\rm (H)}]$,
$d_1^{\rm (H)}\eq d_{\rm opt}^{(q)}\,[n,n-u_1]$,
$2d_1^{\rm (H)}\leq (s_1+2)d_0^{\rm (H)}$,
$\cV_0$ is the $[m,m-1,2]$ parity code over
$(GF(q))^{n-u_1}$, $\cV_1$ is an\\ $[m,s_0,s_1+2]$ MDS code over
$(GF(q))^{u_1-u_0}$ and $d_{\rm opt}^{(q)}\,[(s_1+1)n,s_1(n-u_1)]\leq 2d_1^{\rm (H)}$.
Then, the minimum distance $d$ of $\C(n,\uu)$ is $d\eq 2d_1^{\rm (H)}$
and it meets bound~(\ref{boundcmd}).
}
\end{lemma}
\noindent\pf
By Theorem~\ref{theo1} and since $m\eq s_0+s_1+1$, $d\geq \min\{2d_1^{\rm (H)}\,,\,(s_1+2)d_0^{\rm
(H)}\}\eq 2d_1^{\rm (H)}$.
Since $k^*\eq n-u_0$ and the code has dimension $k\eq s_0(n-u_0)+s_1(n-u_1)$,
then $s_0+1\leq\lceil k/k^*\rceil\leq s_0+s_1\eq m-1$.

Taking $j\eq s_0$ in~(\ref{boundcmd}), we obtain $d\leq d_{\rm
opt}^{(q)}\,[(s_1+1)n,s_1(n-u_1)]\leq 2d_1^{\rm (H)}$.
\qed

\begin{example}
\label{ex19}
{\em Consider the 2-level EII code
$\C(8,(1,1,4,4,8))$ over $GF(2)$ such that $\C_0$ is an
$[8,7,2]$ parity code, $\C_1$ an $[8,4,4]$ extended Hamming code,
$\cV_1$ a $[5,2,4]$ (shortened) RS code over $GF(8)$ and $\cV_0$ a
$[5,4,2]$ parity code over $(GF(2))^4$. According to
Theorem~\ref{theo1}, the minimum distance of the code satisfies $d\geq 8$.
Since $k^*\eq 7$ and $k\eq 22$, $\lceil k/k^*\rceil\eq 4$. Taking
$j\eq s_0\eq 2$ in bound~(\ref{boundcmd}), $d\leq d_{\rm
opt}^{(2)}\,[24,8]\eq 8$~\cite{gr},
so $d\eq 8$ and the bound is met.
}
\end{example}

\begin{example}
\label{ex19bis}
{\em Consider the 2-level EII code
$\C(16,(9,9,11,16,16))$ over $GF(2)$ such that $\C_0$ is a
$[16,7,6]$ extended BCH code, $\C_1$ a $[16,5,8]$ extended BCH code,
$\cV_1$ a $[5,2,4]$ doubly extended RS code over $GF(4)$ and $\cV_0$ a
$[5,3,3]$ (shortened) RS code over $GF(32)$. According to
Theorem~\ref{theo1}, $k\eq 19$ and the minimum distance of the code
satisfies $d\geq 24$.
Since $k^*\eq 7$, $\lceil k/k^*\rceil\eq 3$. Taking
$j\eq 2$ in bound~(\ref{boundcmd}), $d\leq d_{\rm
opt}^{(2)}\,[48,5]\leq 24$ by the Griesmer bound,
so $d\eq 24$ and the bound is met.
}
\end{example}

\begin{example}
\label{ex31}
{\em Consider the 2-level EII code
$\C(15,(\overbrace{10,10,\ldots,10}^{m-2},12,15))$ over $GF(4)$ such that $\C_0$ is a
$[15,5,8]$ BCH code, $\C_1$ a $[15,3,11]$ BCH code,
$\cV_1$ an $[m,m-2,3]$ (shortened) doubly extended RS code over $GF(16)$
$\cV_0$ an $[m,m-1,2]$ parity code over $(GF(2))^3$ and $m\leq 17$.
According to
Theorem~\ref{theo1}, $k\eq 5(m-2)+3$ and the minimum distance of the
code satisfies $d\geq 22$.
Since $k^*\eq 5$ and $\lceil k/k^*\rceil\eq m-1$. Taking
$j\eq m-2$ in bound~(\ref{boundcmd}), $d\leq d_{\rm
opt}^{(4)}\,[30,3]\leq 22$ by the Griesmer bound,
so $d\eq 22$ and the bound is met.
}
\end{example}

Let us compare next EII codes with $s_t>0$ with II codes (i.e.,
$s_t\eq 0$). Constructions of binary $t$-level II codes were presented
in~\cite{hyus,hys}, where an equivalent definition of $t$-level II
codes using generalized tensor product codes is given. Let us
concentrate first on code $\C_{\rm II}$ in~\cite{hyus}. This
construction 
corresponds to an II-code $\C(2^b,\uu)$,
where
$$\uu\eq \left(\overbrace{1,1,\ldots,1}^{m-(\mu-1)}\,,\,b+1\,,\,b
\left(\left\lceil\frac{\mu}{\mu-2}\right\rceil-1\right)+1\,,\,b
\left(\left\lceil\frac{\mu}{\mu-3}\right\rceil-1\right)+1\,,\,\ldots\,,\,b
(\mu-1)+1\right),$$
$m\leq 2^b+1$, the horizontal codes are extended BCH codes
$\C'_{\mu-1}\subseteq\C'_{\mu-2}\subseteq\cdots\subseteq\C'_{1}\subset\C_0$
(some of them can be repeated),
where $\C'_i$ is a $[2^b,2^b-\left(\left(\lceil \mu/\mu-i\rceil-1\right)b+1\right),2\lceil
\mu/\mu-i\rceil]$ code for $0\leq i\leq \mu-1$, the vertical codes are (shortened)
doubly extended RS codes $\cV'_i$, where $\cV'_i$ is an
$[m,m-\left(\lceil\mu/\mu-i\rceil-1\right),2\lceil\mu/\mu-i\rceil]$ over
$GF(2^b)$ for $1\leq i\leq \mu-1$,
$\cV'_0$ is the whole space
$(GF(2^{2^b-(b(\mu-1)+1)})^m$) and $\mu\leq m\leq 2^b+1$.  The
minimum distance of $\C_{\rm II}$ is then, by
Theorem~\ref{theo1}, $d\geq 2\mu$ (notice that the given distances
of BCH codes are designed distances, they may be smaller than the
actual minimum distances~\cite{ms}). For example, if $b\eq 5$ and
$\mu\eq 5$, then $\C_{\rm II}$ is the 3-level II code
$\C(32,(\overbrace{1,1,\ldots,1}^{m-4}\,,\,6\,,\,6\,,\,11\,,\,21))$,
where $\C'_0$ is a $[32,31,2]$ code, $\C'_1\eq\C'_2$ are $[32,26,4]$
codes, $\C'_3$ is a $[32,21,6]$ code and $\C'_4$ is a $[32,11,10]$
code, all these codes nested extended BCH codes, while $\cV'_0$ is
the whole space $\left(GF(2^{11})\right)^m$, $\cV'_1\eq\cV'_2$ is an
$[m,m-1,2]$ code, $\cV'_3$ is an $[m,m-2,3]$ code and $\cV'_4$ is an
$[m,m-4,5]$ code, all these codes (shortened) doubly extended RS
codes over $GF(32)$ and $5\leq m\leq 33$.

Construction $\C_{\rm II}$ in~\cite{hyus} cannot handle cases in
which $2\mu>2^b$, while this is not the case for $t$-level EII codes
with $s_t>0$.
Let us illustrate the possible advantages of EII codes with an example.

\begin{example}
\label{ex32}
{\em Take for instance a $\C_{\rm II}$ code as in~\cite{hyus} with $b\eq 5$
and $\mu\eq 6$, then the minimum distance is $d\geq 12$ and $6\leq
m\leq 33$. The locality
of this code, that is, the number of bits necessary to recover an
erased bit, is $r\eq 2^b-1\eq 31$. Regarded as an II code, it is a
$4$-level
$\C(32,(\overbrace{1,1,\ldots,1}^{m-5}\,,\,6\,,\,6\,,\,6\,,\,11\,,\,26))$
code. Take for instance $m\eq 8$, then it is a
$\C(32,(1\,,\,1\,,\,1\,,\,6\,,\,6\,,\,6\,,\,11\,,\,26))$ II code. As a
binary code, it is a $[256,198,12]$ LRC code with locality 31.

Next, consider a binary 3-level EII code
$\C(16,(\overbrace{1,1,\ldots,1}^{11}\,,\,5\,,\,5\,,\,5\,,\,9\,,\,16))$,
where $\C_0$ is a $[16,15,2]$ code, $\C_1$ is a $[16,11,4]$ code and
$\C_2$ is a $[16,7,6]$ code, all these codes nested extended BCH
codes, while $\cV_0$ is a $[16,15,2]$ parity code over $(GF(2))^7$,
$\cV_1$ is a $[16,14,3]$ code and $\cV_2$ is a $[16,11,6]$ code, both
these codes extended RS codes over $GF(16)$. By Theorem~\ref{theo1},
the minimum distance of this code is $d\geq 12$. As a binary code, it
is a $[256,205,12]$ LRC code with locality 15, so it has better rate
than the $\C_{\rm II}$ code and reduces the locality in half.
}
\end{example}

In order to compare with the codes in~\cite{hys}, we go back to the
conditions of Corollary~\ref{cor5}. The next example shows that the
improvement in minimum distance for the same parameters may be significant.

\begin{example}
\label{ex25}
{\em
Consider a $4$-level EII code $\C(9,\uu)$ as given by
Definition~\ref{defEII}, where $q\eq 8$,

\begin{eqnarray*}
\uu &=&
\left(\overbrace{1,1,\ldots,1}^{s_0},\overbrace{3,3,\ldots,3}^{s_1},\overbrace{5,5,\ldots,5}^{s_2},
\overbrace{7,7,\ldots,7}^{s_3},\overbrace{9,9,\ldots,9}^{s_4}\right),
\end{eqnarray*}
$m\eq s_0+s_1+s_2+s_3+s_4$, $s_4\geq 0$,
$\C_3\subset\C_2\subset\C_1\subset\C_0$ are doubly extended RS codes
over $GF(8)$ such that
$\C_0$ is a $[9,8,2]$ code, $\C_1$ is a $[9,6,4]$ code, $\C_2$
is a $[9,4,6]$ code and $\C_3$ is a $[9,2,8]$ code,
$\cV_0,\cV_1,\cV_2,\cV_3$ are doubly extended RS codes
over $GF(64)$ such that
$\cV_0$ is an $[m,m-\hs_4,\hs_4+1]$ code, $\cV_1$ is an $[m,m-\hs_3,\hs_3+1]$ code,
$\cV_2$ is an $[m,m-\hs_2,\hs_2+1]$ code and $\cV_3$ is an
$[m,m-\hs_1,\hs_1+1]$ code. We can see that in this case, the $x$
defined after Corollary~\ref{cor5} is $x\eq 2$, so $m\leq 8^2+1\eq 65$. Let us take $m\eq
65$. By Corollary~\ref{cor5}, $\C(9,\uu)$ has dimension
$585-s_0-3s_1-5s_2-7s_3-9s_4$ and minimum distance

\begin{eqnarray}
\label{eq9}
d&\geq &\min\{2(\hs_{1}+1),4(\hs_{2}+1),6(\hs_{3}+1),8(\hs_{4}+1)\}.
\end{eqnarray}

Assume that $\C(9,\uu)$ is an II code, i.e., $s_4\eq 0$, the
situation studied in~\cite{hys} with generalized tensor product
codes. In this case we obtain $d\eq 8$.


If we take an EII code that is not an II code (i.e., $s_4>0$) with
the same rate, we can improve upon the minimum distance of 8.
For example, consider an II code with $s_0\eq 17$ and $s_1\eq s_2\eq s_3\eq
16$, and an EII code with $s_0\eq s_1\eq 20$, $s_2\eq 10$, $s_3\eq 4$
and $s_4\eq 11$. Both codes have 257 parity symbols and hence the
same rate, but the II code, as we have seen, has minimum distance
$d\eq 8$, while the EII code, by~(\ref{eq9}), has minimum distance
$d\geq \min\{(2)(46),(4)(26),(6)(16),(8)(12)\}\eq 92$.
}
\end{example}

We take a final example.

\begin{example}
\label{ex25bis}
{\em
Consider a $3$-level II code $\C^{\rm (I)}\eq\C(16,(\overbrace{2,2,\ldots,2}^{254},4,8))$ as given by
Definition~\ref{defEII}, where $q\eq 16$,
$\C_2\subset\C_1\subset\C_0$ are extended RS codes
over $GF(16)$ such that
$\C_0$ is a $[16,14,3]$ code, $\C_1$ is a $[16,12,5]$ code, $\C_2$ is a $[16,8,9]$ code,
$\cV_1$ is a $[256,255,2]$ code over $(GF(16))^8$ and
$\cV_2$ is a $[256,254,3]$ code over $GF(256)$. By
Corollary~\ref{cor5}, 
$\C^{\rm (I)}$ has dimension $3576$ and minimum distance 9, i.e., it is a
$[4096,3576,9]$ code over $GF(16)$. In
particular, it is a special case of Example~4 in~\cite{hys}, where,
following the notation in that example, $\mu\eq 3$, $d'_3\eq 9$,
$d'_2\eq 5$, $d'_1\eq 3$, $\delta_3\eq 2$ and $\delta_2\eq 3$. It is
proven in Theorem~9 of~\cite{hys} that $\C^{\rm (I)}$  has the largest
possible dimension among all codes with the same erasure-correcting
capability.

Consider next a $3$-level EII code
$\C^{\rm (II)}\eq
\C(16,(\overbrace{1,1,\ldots,1}^{216},\overbrace{3,3,\ldots,3}^{20},\overbrace{5,5,\ldots,5}^{7},\overbrace{16,16,\ldots,16}^{13}))$
as given by
Definition~\ref{defEII}, where also $q\eq 16$,
$\C_2\subset\C_1\subset\C_0$ are extended RS codes
over $GF(16)$ such that
$\C_0$ is a $[16,15,2]$ code, $\C_1$ is a $[16,13,4]$ code, $\C_2$
is a $[16,11,6]$ code, $\cV_0$ is a $[256,243,14]$ code over $(GF(16))^{11}$,
$\cV_1$ is a $[256,236,21]$ code over $GF(256)$ and
$\cV_2$ is a $[256,216,41]$ code over $GF(256)$. By
Corollary~\ref{cor5}, $\C^{\rm (II)}$
has dimension
$3577$ and minimum distance 82, i.e., it is a $[4096,3577,82]$ code
over $GF(16)$ whose minimum distance is significantly larger than the
one of $\C^{\rm (I)}$ and its dimension is also larger.

As done in Example~\ref{ex2}, we may consider the average number of
erasures that both codes can correct. Doing a Montecarlo simulation,
we find out that $\C^{\rm (I)}$    
can correct an average of 119 erasures, while $\C^{\rm (II)}$
can correct an average of 184 erasures. 

Consider next the 4-level II code
$\C^{\rm
(III)}\eq\C(16,(\overbrace{1,1,\ldots,1}^{143},\overbrace{3,3,\ldots,3}^{99},\overbrace{5,5,\ldots,5}^{9},\overbrace{7,7,\ldots,7}^{5}))$,
$\C_0$, $\C_1$ and $\C_2$ are like in $\C^{\rm (II)}$,
$\C_3\subset\C_2$ is a
$[16,9,8]$ code over $GF(16)$, $\cV_1$
is a $[256,251,6]$ code, $\cV_2$ is a $[256,242,15]$ code and $\cV_3$
is a $[256,143,114]$ code, all three codes over $GF(256)$.
As $\C^{\rm (I)}$, $\C^{\rm (III)}$ is a $[4096,3576,8]$ code over $GF(16)$ by Corollary~\ref{cor5}, hence, its
minimum distance is smaller than the one of $\C^{\rm (I)}$ and considerably smaller than the one of $\C^{\rm (II)}$.
However, the average number of erasures it can
correct is 369, much larger than the average of 184 erasures that $\C^{\rm (II)}$ can correct.
}
\end{example}

\section{Systematic Encoding and Decoding of EII Codes}
\label{encoding}
Definition~\ref{defEII} implicitly gives a simple encoding algorithm. However,
such algorithm is not systematic. Next we present a systematic encoding algorithm.

\begin{algorithm}[Systematic Encoding Algorithm]
\label{algencodingsyst}
{\em Consider a $t$-level EII code $\C(n,\uu)$ according to
Definition~\ref{defEII}. Assume that the data is given by $D_{i,j}$,
where, for each $v$ such that $0\leq v\leq t-1$, $0\leq i\leq
(m-\hs_{v+1})-1$ and $n-u_{v+1}\leq j\leq n-u_v-1$.
We proceed by induction on $t$.

If $t\eq 1$ then the data $D_{i,j}$, where $0\leq i\leq m-s_1-1\eq
s_0-1$, $n-u_1\leq j\leq n-u_0-1$,
is encoded into the (vertical) $[m,s_0,d_0^{\rm (V)}]$ code
$\cV_0$ over $(GF(q))^{u_1-u_0}$, and an $m\times (n-u_1)$ zero array
is appended to this vertically encoded array from the left to obtain
an $m\times (n-u_0)$ array. Then each of the $m$ rows of this array
is encoded systematically into the (horizontal) code $\C_0$ to give
the final $m\times n$ encoded array.

Next, assume that $t>1$ and by induction, assume that there is a systematic encoder
for any $(t-1)$-level EII code. Consider the $(t-1)$-level EII code
$\C(n,\uu')$, where $\uu'$ is given by~(\ref{equu'}),
and encode systematically the
data $D_{i,j}$ for $1\leq v\leq t-1$, $0\leq i\leq (m-\hs_{v+1})-1$ and
$n-u_{v}\leq j\leq n-u_{v-1}-1$ into an array $C'\in\C(n,\uu')$. Denote by
$P'_{i,j}$ the parity symbols obtained as a result of this systematic
encoder. Next, encode
systematically the symbols $D_{i,j}\xor P'_{i,j}$ for
$0\leq i\leq s_0-1$ and $n-u_1\leq j\leq n-u_0-1$ into an array $C^{(0)}$ in
the 1-level EII code $\C(n,\uu^{(0)})$, where $\uu^{(0)}$ is given by~(\ref{equui}).
Then, the final
encoded array is 
$C\eq C^{(0)}\,\xor\, C'$.
\qed
}
\end{algorithm}

\vspace{-3mm}

\begin{example}
\label{ex6}
{\em
Consider the $\C(15,(4,4,8,10,15))$ 3-level EII code over $GF(2)$ of
Example~\ref{ex4} and assume that the data we want to encode is the following:

$$
\begin{array}{|c|c|c|c|c||c|c||c|c|c|c|}
\hline
D_{0,0}&D_{0,1}&D_{0,2}&D_{0,3}&D_{0,4}&D_{0,5}&D_{0,6}&D_{0,7}&D_{0,8}&D_{0,9}&D_{0,10}\\
\hline
D_{1,0}&D_{1,1}&D_{1,2}&D_{1,3}&D_{1,4}&D_{1,5}&D_{1,6}&D_{1,7}&D_{1,8}&D_{1,9}&D_{1,10}\\
\hline
D_{2,0}&D_{2,1}&D_{2,2}&D_{2,3}&D_{2,4}&D_{2,5}&D_{2,6}\\
\cline{1-7}
D_{3,0}&D_{3,1}&D_{3,2}&D_{3,3}&D_{3,4}\\
\cline{1-5}
\end{array}
$$

We will show how to encode in systematic form this
data by applying iteratively Algorithm~\ref{algencodingsyst}.

First we encode the data $D_{i,j}$ for $0\leq i\leq 3$ and $0\leq j\leq 4$
into the 1-level EII code $\C(15,(10,10,10,10,15))$ to obtain $C^{(2)}$ as follows:

\begin{eqnarray*}
C^{(2)}&=&
\begin{array}{|c|c|c|c|c||c|c|c|c|c|c|c|c|c|c|}
\hline
D_{0,0}&D_{0,1}&D_{0,2}&D_{0,3}&D_{0,4}&P^{(2)}_{0,5}&P^{(2)}_{0,6}&P^{(2)}_{0,7}&P^{(2)}_{0,8}&P^{(2)}_{0,9}&P^{(2)}_{0,10}&P^{(2)}_{0,11}&P^{(2)}_{0,12}&P^{(2)}_{0,13}&P^{(2)}_{0,14}\\
\hline
D_{1,0}&D_{1,1}&D_{1,2}&D_{1,3}&D_{1,4}&P^{(2)}_{1,5}&P^{(2)}_{1,6}&P^{(2)}_{1,7}&P^{(2)}_{1,8}&P^{(2)}_{1,9}&P^{(2)}_{1,10}&P^{(2)}_{1,11}&P^{(2)}_{1,12}&P^{(2)}_{1,13}&P^{(2)}_{1,14}\\
\hline
D_{2,0}&D_{2,1}&D_{2,2}&D_{2,3}&D_{2,4}&P^{(2)}_{2,5}&P^{(2)}_{2,6}&P^{(2)}_{2,7}&P^{(2)}_{2,8}&P^{(2)}_{2,9}&P^{(2)}_{2,10}&P^{(2)}_{2,11}&P^{(2)}_{2,12}&P^{(2)}_{2,13}&P^{(2)}_{2,14}\\
\hline
D_{3,0}&D_{3,1}&D_{3,2}&D_{3,3}&D_{3,4}&P^{(2)}_{3,5}&P^{(2)}_{3,6}&P^{(2)}_{3,7}&P^{(2)}_{3,8}&P^{(2)}_{3,9}&P^{(2)}_{3,10}&P^{(2)}_{3,11}&P^{(2)}_{3,12}&P^{(2)}_{3,13}&P^{(2)}_{3,14}\\
\hline\hline
P^{(2)}_{4,0}&P^{(2)}_{4,1}&P^{(2)}_{4,2}&P^{(2)}_{4,3}&P^{(2)}_{4,4}&P^{(2)}_{4,5}&P^{(2)}_{4,6}&P^{(2)}_{4,7}&P^{(2)}_{4,8}&P^{(2)}_{4,9}&P^{(2)}_{4,10}&P^{(2)}_{4,11}&P^{(2)}_{4,12}&P^{(2)}_{4,13}&P^{(2)}_{4,14}\\
\hline
\end{array}
\end{eqnarray*}

Encoding $D^{(1)}_{i,j}\eq D_{i,j}\xor P^{(2)}_{i,j}$ for $0\leq i\leq 2$ and $5\leq j\leq 6$
into $\C(15,(8,8,8,10,10))$, we obtain

\begin{eqnarray*}
C^{(1)}&=&
\begin{array}{|c|c|c|c|c||c|c||c|c|c|c|c|c|c|c|}
\hline
\hspace{2mm}0\hspace{2mm}&\hspace{2mm}0\hspace{2mm}&\hspace{2mm}0\hspace{2mm}&\hspace{2mm}0\hspace{2mm}&\hspace{2mm}0\hspace{2mm}&
D^{(1)}_{0,5}&D^{(1)}_{0,6}&P^{(1)}_{0,7}&P^{(1)}_{0,8}&P^{(1)}_{0,9}&P^{(1)}_{0,10}&P^{(1)}_{0,11}&P^{(1)}_{0,12}&P^{(1)}_{0,13}&P^{(1)}_{0,14}\\
\hline
0&0&0&0&0&D^{(1)}_{1,5}&D^{(1)}_{1,6}&P^{(1)}_{1,7}&P^{(1)}_{1,8}&P^{(1)}_{1,9}&P^{(1)}_{1,10}&P^{(1)}_{1,11}&P^{(1)}_{1,12}&P^{(1)}_{1,13}&P^{(1)}_{1,14}\\
\hline
0&0&0&0&0&D^{(1)}_{2,5}&D^{(1)}_{2,6}&P^{(1)}_{2,7}&P^{(1)}_{2,8}&P^{(1)}_{2,9}&P^{(1)}_{2,10}&P^{(1)}_{2,11}&P^{(1)}_{2,12}&P^{(1)}_{2,13}&P^{(1)}_{2,14}\\
\hline\hline
0&0&0&0&0&P^{(1)}_{3,5}&P^{(1)}_{3,6}&P^{(1)}_{3,7}&P^{(1)}_{3,8}&P^{(1)}_{3,9}&P^{(1)}_{3,10}&P^{(1)}_{3,11}&P^{(1)}_{3,12}&P^{(1)}_{3,13}&P^{(1)}_{3,14}\\
\hline
0&0&0&0&0&P^{(1)}_{4,5}&P^{(1)}_{4,6}&P^{(1)}_{4,7}&P^{(1)}_{4,8}&P^{(1)}_{4,9}&P^{(1)}_{4,10}&P^{(1)}_{4,11}&P^{(1)}_{4,12}&P^{(1)}_{4,13}&P^{(1)}_{4,14}\\
\hline
\end{array}
\end{eqnarray*}

Encoding $D^{(0)}_{i,j}\eq D_{i,j}\xor P^{(1)}_{i,j}\xor P^{(2)}_{i,j}$ for $0\leq i\leq 1$ and $7\leq j\leq 10$
into $\C(15,(4,4,8,8,8))$, we obtain

\begin{eqnarray*}
C^{(0)}&=&
\begin{array}{|c|c|c|c|c|c|c||c|c|c|c||c|c|c|c|}
\hline
\hspace{2mm}0\hspace{2mm}&\hspace{2mm}0\hspace{2mm}&\hspace{2mm}0\hspace{2mm}&\hspace{2mm}0\hspace{2mm}&\hspace{2mm}0\hspace{2mm}&
\hspace{2mm}0\hspace{2mm}&\hspace{2mm}0\hspace{2mm}&D^{(0)}_{0,7}&D^{(0)}_{0,8}&D^{(0)}_{0,9}&D^{(0)}_{0,10}&P^{(0)}_{0,11}&P^{(0)}_{0,12}&P^{(0)}_{0,13}&P^{(0)}_{0,14}\\
\hline
0&0&0&0&0&0&0&D^{(0)}_{1,7}&D^{(0)}_{1,8}&D^{(0)}_{1,9}&D^{(0)}_{1,10}&P^{(0)}_{1,11}&P^{(0)}_{1,12}&P^{(0)}_{1,13}&P^{(0)}_{1,14}\\
\hline\hline
0&0&0&0&0&0&0&P^{(0)}_{2,7}&P^{(0)}_{2,8}&P^{(0)}_{2,9}&P^{(0)}_{2,10}&P^{(0)}_{2,11}&P^{(0)}_{2,12}&P^{(0)}_{2,13}&P^{(0)}_{2,14}\\
\hline
0&0&0&0&0&0&0&P^{(0)}_{3,7}&P^{(0)}_{3,8}&P^{(0)}_{3,9}&P^{(0)}_{3,10}&P^{(0)}_{3,11}&P^{(0)}_{3,12}&P^{(0)}_{3,13}&P^{(0)}_{3,14}\\
\hline
0&0&0&0&0&0&0&P^{(0)}_{4,7}&P^{(0)}_{4,8}&P^{(0)}_{4,9}&P^{(0)}_{4,10}&P^{(0)}_{4,11}&P^{(0)}_{4,12}&P^{(0)}_{4,13}&P^{(0)}_{4,14}\\
\hline
\end{array}
\end{eqnarray*}

The encoded array is then $C\eq C^{(2)}\xor C^{(1)}\xor C^{(0)}$, which gives the
following array in systematic form:

$$
\begin{array}{|c|c|c|c|c|c|c|c|c|c|c|c|c|c|c|}
\hline
D_{0,0}&D_{0,1}&D_{0,2}&D_{0,3}&D_{0,4}&D_{0,5}&D_{0,6}&D_{0,7}&D_{0,8}&D_{0,9}&D_{0,10}&P_{0,11}&P_{0,12}&P_{0,13}&P_{0,14}\\
\hline
D_{1,0}&D_{1,1}&D_{1,2}&D_{1,3}&D_{1,4}&D_{1,5}&D_{1,6}&D_{1,7}&D_{1,8}&D_{1,9}&D_{1,10}&P_{1,11}&P_{1,12}&P_{1,13}&P_{1,14}\\
\hline
D_{2,0}&D_{2,1}&D_{2,2}&D_{2,3}&D_{2,4}&D_{2,5}&D_{2,6}&P_{2,7}&P_{2,8}&P_{2,9}&P_{2,10}&P_{2,11}&P_{2,12}&P_{2,13}&P_{2,14}\\
\hline
D_{3,0}&D_{3,1}&D_{3,2}&D_{3,3}&D_{3,4}&P_{3,5}&P_{3,6}&P_{3,7}&P_{3,8}&P_{3,9}&P_{3,10}&P_{3,11}&P_{3,12}&P_{3,13}&P_{3,14}\\
\hline
P_{4,0}&P_{4,1}&P_{4,2}&P_{4,3}&P_{4,4}&P_{4,5}&P_{4,6}&P_{4,7}&P_{4,8}&P_{4,9}&P_{4,10}&P_{4,11}&P_{4,12}&P_{4,13}&P_{4,14}\\
\hline
\end{array}
$$
}
\end{example}

Before stating the
error-erasure correcting capability of $t$-level EII codes, we give a lemma.

\begin{lemma}
\label{l3}
{\em Consider an array $C$ in a $t$-level EII code $\C(n,\uu)$ as
given by Definition~\ref{defEII}, hence,
$C\eq\bigoplus_{v=0}^{t-1}C^{(v)}$, where\\ $C^{(v)}\in\C(n,\uu^{(v)})$
for $0\leq v\leq t-1$ and $\uu^{(v)}$ is given by~(\ref{equui}).
Denote by $\uc_i$ each row of $C$ and by
$\uc^{(v)}_i$ each row of $C^{(v)}$, where $0\leq i\leq m-1$, thus,
$\uc_i\eq \bigoplus_{v=0}^{t-1}\uc^{(v)}_i$. Then, given $\uc_i$,
each $\uc^{(v)}_i$ can be obtained for $0\leq v\leq t-1$.
}
\end{lemma}
\pf We do induction on $t$. If $t\eq 1$, then $C\eq C^{(0)}$ and the
result follows since $\uc_i\eq\uc^{(0)}_i$, so assume that $t>1$.

Since $c_{i,j}\eq
c^{(t-1)}_{i,j}$ for $0\leq j\leq n-u_{t-1}-1$, then encoding
(systematically) $(c^{(t-1)}_{i,0},c^{(t-1)}_{i,1},\ldots
,c^{(t-1)}_{i,n-u_{t-1}-1})$ using code $\C_{t-1}$, we obtain
$(c^{(t-1)}_{i,0},c^{(t-1)}_{i,1},\ldots,c^{(t-1)}_{i,n-1})\eq
\uc^{(t-1)}_{i}$. Then,
$\uc_i\xor\uc^{(t-1)}_{i}\eq\bigoplus_{v=0}^{t-2}\uc^{(v)}_i$. Notice
that $\bigoplus_{v=0}^{t-2}\uc^{(v)}_i$ is the $i$th row of
$C'\eq\bigoplus_{v=0}^{t-2}C^{(v)}$. But $C'$, by
Definition~\ref{defEII}, is in a $(t-1)$-level
EII code,
so, by induction, we can obtain each  $\uc^{(v)}_i$ for $0\leq v\leq t-2$.
\qed

\begin{example}
\label{ex7}
{\em
Consider the $\C(15,(4,4,8,10,15))$ 3-level EII code over $GF(2)$ of
Examples~\ref{ex4} and~\ref{ex6}.
Assume that\\ $C\in\C(15,(4,4,8,10,15))$, where
$C\eq (c_{i,j})_{\substack{0\leq i\leq 4\\0\leq j\leq 14}}$.
Then $C\eq C^{(2)}\xor C^{(1)}\xor C^{(0)}$, where
$C^{(2)}\in\C(15,(10,10,10,10,15))$, $C^{(1)}\in\C(15,(8,8,8,10,10))$
and $C^{(0)}\in\C(15,(4,4,8,8,8))$. Assume that a row $\uc_i\eq
(c_{i,0},c_{i,1},\ldots,c_{i,14})$ is given and we want to obtain the
rows $\uc^{(v)}_i$, where $0\leq i\leq 4$ and $0\leq v\leq 2$. Since
$c_{i,j}\eq c^{(2)}_{i,j}$ for $0\leq j\leq 4$, we encode
(systematically) $(c^{(2)}_{i,0},c^{(2)}_{i,1},\ldots,c^{(2)}_{i,4})$
in code $\C_2$ and we obtain codeword $\uc^{(2)}_i$. Next we compute
$\uc_i\xor\uc^{(2)}_i\eq\uc^{(1)}_i\xor\uc^{(0)}_i$. Notice that
$c^{(0)}_{i,j}\eq 0$ for $0\leq j\leq 6$, so, $c^{(1)}_{i,j}\eq
c_{i,j}\xor c^{(2)}_{i,j}$ for $0\leq j\leq 6$. Encoding systematically
$(c^{(1)}_{i,0},c^{(1)}_{i,1},\ldots,c^{(1)}_{i,6})$
in code $\C_1$ we obtain codeword $\uc^{(1)}_i$. Finally, codeword
$\uc^{(0)}_i$ is obtained as $\uc^{(0)}_i\eq \uc\xor\uc^{(2)}_i\xor\uc^{(1)}_i$.
}
\end{example}


Next we give the error-erasure correction of the codes under the
assumption that there are no miscorrections.

\begin{theorem}
\label{theo3}
{\em
Consider a $t$-level EII code $\C(n,\uu)$ as given by
Definition~\ref{defEII}. Let $C\in\C(n,\uu)$. Then $\C(n,\uu)$ can
correct in $C$ any row with $x_0$ errors and $y_0$ erasures, where
$2x_0+y_0\leq d^{\rm (H)}_{0}-1$, any up
to $d^{\rm (V)}_{i}-d^{\rm (V)}_{i-1}$ rows
with $x_{t-i}$ errors and $y_{t-i}$ erasures each, where
$2x_{t-i}+y_{t-i}\leq d^{\rm (H)}_{t-i}-1$ for $1\leq i\leq t-1$
and any up to $d^{\rm (V)}_{0}-1$ rows with $x_{t}$
errors and $y_{t}$ erasures each, where $2x_{t}+y_{t}\geq d^{\rm
(H)}_{t-1}$, provided that when correction of a row in the array is
attempted, the decoder will either correctly decode such row  or
it will declare an uncorrectable error (i.e., there are no
miscorrections).
}
\end{theorem}
\pf Let $C\in\C(n,\uu)$
having up to $d^{\rm (V)}_{i}-d^{\rm (V)}_{i-1}$ rows
with $x_{t-i}$ errors and $y_{t-i}$ erasures each, where
$2x_{t-i}+y_{t-i}\leq d^{\rm (H)}_{t-i}-1$ for $1\leq i\leq t-1$
and up to $d^{\rm (V)}_{0}-1$ rows with $x_{t}$
errors and $y_{t}$ erasures each, where $2x_{t}+y_{t}\geq d^{\rm
(H)}_{t-1}$, while the remaining rows have $x_0$ errors and
$y_0$ erasures, where $2x_0+y_0\leq d^{\rm (H)}_{0}-1$.

Since each row in the array is in $\C_0$, the rows in $C$ with
$x_0$ errors and $y_0$ erasures, where
$2x_0+y_0\leq d^{\rm (H)}_{0}-1$,
are corrected, while, given the theorem's
assumption, the remaining rows are detected as having an
uncorrectable pattern (i.e., no miscorrection).

Denote the rows 
detected as uncorrectable
by $\uc_{i_0},\uc_{i_1},\ldots,\uc_{i_{\ell-1}}$,
where
\begin{eqnarray*}
\ell\;\leq\;\left(\sum_{i=1}^{t-1}(d^{\rm (V)}_{i}-d^{\rm
(V)}_{i-1})\right)+d^{\rm (V)}_{0}-1\;\eq\; d^{\rm
(V)}_{t-1}-1,
\end{eqnarray*}
and the remaining rows (which are error and erasure free) by
$\uc_{j_0},\uc_{j_1},\ldots,\uc_{j_{m-\ell-1}}$.
We do induction on $t$.

Assume that $t\eq 1$. Then there are $\ell$ rows with $x_1$
errors and $y_1$
erasures each, where $\ell\leq d^{\rm (V)}_{0}-1$
and $2x_1+y_1\geq d^{\rm (H)}_{0}$.
The vectors $(c_{j_w,n-u_1},c_{j_w,n-u_1+1},\ldots,c_{j_w,n-u_0-1})$, where
$0\leq w\leq m-\ell-1$,
are entries in a codeword in the vertical code $\cV_0$. Since $\cV_0$
has minimum distance $d^{\rm (V)}_{0}$, the vectors
$(c_{i_w,n-u_1},c_{i_w,n-u_1+1},\ldots,c_{i_w,n-u_0-1})$
for $0\leq w\leq \ell-1$ can be recovered by doing (vertical) erasure
decoding. Once these vectors are recovered, for each $w$ such that
$0\leq w\leq \ell-1$, we encode systematically each
$(c_{i_w,0},c_{i_w,1},\ldots,c_{i_w,n-u_0-1})$ into
$\uc_{i_w}\in\C_0$, completing the decoding (notice that
$c_{i_w,j}\eq 0$ for $0\leq j\leq n-u_1-1$).

Next assume that $t>1$. For each $w$ such that $0\leq w\leq
m-\ell-1$, we have seen that the rows $\uc_{j_w}$ in $C$ are erasure
free. Since
$\uc_{j_w}\eq\bigoplus_{v=0}^{t-1}\uc^{(v)}_{j_w}$, where
$\uc^{(v)}_{j_w}\in\C_v$, by Lemma~\ref{l3}, we can obtain
$\uc^{(v)}_{j_w}\in\C_v$ for each $v$ such that $0\leq v\leq t-1$ and
for each $w$ such that $0\leq w\leq m-\ell-1$. In particular, we obtain
$\uc^{(0)}_{j_w}\in\C_0$. Using the vectors
$(c^{(0)}_{j_w,n-u_1},c^{(0)}_{j_w,n-u_1+1},\ldots,c^{(0)}_{j_w,n-u_0-1})$, where
$0\leq w\leq m-\ell-1$, we retrieve
$C^{(0)}\in\C(n,\uu^{(0)})$
as in the case of $t\eq 1$ above. Since by Lemma~\ref{cor0},
$C'\eq C\xor C^{(0)}$, where $C'\in\C(n,\uu')$ and $\uu'$ is given
by~(\ref{equu'}), and since $\C(n,\uu')$ is a $(t-1)$-level EII code and
each row is in $\C_1$, the rows with $x_1$ errors and $y_1$
erasures with $2x_1+y_1\leq d^{\rm (V)}_{1}-1$
in $C'$ are corrected and the rows with $x$ errors and $y$ erasures
such that $2x+y\geq d^{\rm (V)}_{1}$ are detected. We are left with up
to $d^{\rm (V)}_{i}-d^{\rm (V)}_{i-1}$ rows
with $x_{t-i}$ errors and $y_{t-i}$ erasures each, where
$2x_{t-i}+y_{t-i}\leq d^{\rm (H)}_{t-i}-1$  for $1\leq i\leq t-2$,
and up to $d^{\rm (V)}_{0}-1$ rows with $x_{t}$ errors
and $y_{t}$ erasures each for $2x_{t}+y_{t}\geq d^{\rm
(H)}_{t-1}$ in $C'$. Then, by induction, these errors and erasures
can be corrected, so $C$ is retrieved as $C\eq C'\xor C^{(0)}$.
\qed

Implicit in the proof of Theorem~\ref{theo3} is a decoding algorithm
which is somewhat similar to the coset decoding method for II codes
given in~\cite{cw}.

\begin{example}
\label{ex10}
{\em
Consider a 3-level EII-code $\C(15,(4,8,10,15))$ over $GF(16)$,
where
$\C_0$, $\C_1$ and $\C_2$ are $[15,11,5]$, $[15,7,9]$ and $[15,5,11]$
RS codes over $GF(16)$ respectively, $\C_2\subset\C_1\subset\C_0$,
$\cV_0$ is a $[4,3,2]$ parity code over $(GF(16))^5$, $\cV_1$ is a
$[4,2,3]$ MDS code over $(GF(16))^2$ and $\cV_2$ is a $[4,1,4]$
repetition code over $(GF(16))^4$.

According to Theorem~\ref{theo3}, given an array in $\C(15,(4,8,10,15))$ with
errors and erasures, this code can correct any row with $x_0$
errors and $y_0$ erasures each, where $2x_0+y_0\leq 4$,
up to one row with $x_1$ errors and $y_1$ erasures, where
$2x_1+y_1\leq 8$,  up to one row with $x_2$ errors and $y_2$ erasures, where
$2x_2+y_2\leq 10$, and up to one row with $x_3$ errors and
$y_3$ erasures, where $2x_3+y_3\geq 11$, provided that there is no
miscorrection each time correction of a row is attempted. For example,
assume that an array was stored, and the following array is received,
where X denotes error and E erasure:

\begin{center}
\begin{tabular}{c|c|c|c|c|c|c|c|c|c|c|c|c|c|c|c|}
\multicolumn{1}{c}{\phantom{0}}&\multicolumn{1}{c}{\scriptsize
0}&\multicolumn{1}{c}{\scriptsize 1}&\multicolumn{1}{c}{\scriptsize 2}&
\multicolumn{1}{c}{\scriptsize 3}& \multicolumn{1}{c}{\scriptsize
4}&\multicolumn{1}{c}{\scriptsize 5}&\multicolumn{1}{c}{\scriptsize
6}&\multicolumn{1}{c}{\scriptsize 7}&\multicolumn{1}{c} {\scriptsize 8}&\multicolumn{1}{c}{\scriptsize 9}&\multicolumn{1}{c}{\scriptsize
10}&\multicolumn{1}{c}{\scriptsize 11}&\multicolumn{1}{c}{\scriptsize 12}&\multicolumn{1}{c}{\scriptsize 13}&\multicolumn{1}{c}{\scriptsize 14}\\
\cline{2-16}
$\uc_0$&&&&&X&&&X&&&X&&&X&\\
\cline{2-16}
$\uc_1$&X&&&&X&X&&&X&&&X&&E&\\
\cline{2-16}
$\uc_2$&&X&&&E&&X&&&E&&X&&&X\\
\cline{2-16}
$\uc_3$&&&E&&&&&X&&&&E&&&\\
\cline{2-16}
\end{tabular}
\end{center}

The first step is correcting the error and two erasures in
$\uc_3$, which can be done since each row $\uc_i$ is in $\C_0$, while
rows 0, 1, and 2 we assume are detected as uncorrectable in $\C_0$. So, we
are left with the following array:

\begin{center}
\begin{tabular}{c|c|c|c|c|c|c|c|c|c|c|c|c|c|c|c|}
\multicolumn{1}{c}{\phantom{0}}&\multicolumn{1}{c}{\scriptsize
0}&\multicolumn{1}{c}{\scriptsize 1}&\multicolumn{1}{c}{\scriptsize 2}&
\multicolumn{1}{c}{\scriptsize 3}& \multicolumn{1}{c}{\scriptsize
4}&\multicolumn{1}{c}{\scriptsize 5}&\multicolumn{1}{c}{\scriptsize
6}&\multicolumn{1}{c}{\scriptsize 7}&\multicolumn{1}{c} {\scriptsize 8}&\multicolumn{1}{c}{\scriptsize 9}&\multicolumn{1}{c}{\scriptsize
10}&\multicolumn{1}{c}{\scriptsize 11}&\multicolumn{1}{c}{\scriptsize 12}&\multicolumn{1}{c}{\scriptsize 13}&\multicolumn{1}{c}{\scriptsize 14}\\
\cline{2-16}
$\uc_0$&&&&&X&&&X&&&X&&&X&\\
\cline{2-16}
$\uc_1$&X&&&&X&X&&&X&&&X&&E&\\
\cline{2-16}
$\uc_2$&&X&&&E&&X&&&E&&X&&&X\\
\cline{2-16}
$\uc_3$&&&&&&&&&&&&&&&\\
\cline{2-16}
\end{tabular}
\end{center}

Consider $\uc_3$, which is
now erasure free. We have,
$\uc_3\eq\uc^{(2)}_3\xor\uc^{(1)}_3\xor\uc^{(0)}_3$, where
$\uc^{(v)}_3\in\C_v$ for $0\leq v\leq 2$, and each $\uc^{(v)}_3$ can
be obtained from $\uc_3$ by Lemma~\ref{l3}.

From the vector
$(c^{(0)}_{3,7}),c^{(0)}_{3,8}),c^{(0)}_{3,9}),c^{(0)}_{3,10}))$, we
can retrieve the vectors
$(c^{(0)}_{i,7}),c^{(0)}_{i,8}),c^{(0)}_{i,9}),c^{(0)}_{i,10}))$ for
$0\leq i\leq 2$ using the (vertical) $[4,1,4]$ repetition code
$\cV_2$ over $(GF(16))^4$.

Next, for $i\in\{0,1,2\}$, we encode the vectors of length 11 over $GF(16)$
$(c^{(0)}_{i,0},c^{(0)}_{i,1},\ldots,c^{(0)}_{i,10})$ into
$\uc^{(0)}_i$ in the $[15,11,5]$ RS code $\C_0$, where
$c^{(0)}_{i,j}\eq 0$ for $0\leq j\leq 6$. We then take
$\uc^{(2)}_i\xor\uc^{(1)}_i\eq \uc_i\xor\uc^{(0)}_i$ for $0\leq i\leq
2$. Since $\uc^{(2)}_i\xor\uc^{(1)}_i$ is in the $[15,7,9]$ RS code
$\C_1$, we attempt correction in $\C_1$ of rows 0, 1 and 2. The only
row that is correctable in $\C_1$ is row 0 since it has 4
errors, while we assume that we detect rows 1 and 2 as uncorrectable.
Once we correct $\uc^{(2)}_0\xor\uc^{(1)}_0$, we obtain
$\uc_0\eq (\uc^{(2)}_0\xor\uc^{(1)}_0)\xor\uc^{(0)}_0$.
Again by Lemma~\ref{l3}, we obtain $\uc^{(1)}_0$ and $\uc^{(2)}_0$
from $\uc_0$.
Taking the entries
$(c^{(1)}_{0,5},c^{(1)}_{0,6})$ and $(c^{(1)}_{3,5},c^{(1)}_{3,6})$
in $(GF(16))^2$, we retrieve the entries $(c^{(1)}_{1,5},c^{(1)}_{1,6})$
and $(c^{(1)}_{2,5},c^{(1)}_{2,6})$ using the $[4,2,3]$ MDS code
$\cV_1$ over $(GF(16))^2$. Then, from
$(c^{(1)}_{i,0},c^{(1)}_{i,1},\ldots,c^{(1)}_{i,6})$, where
$c^{(1)}_{i,j}\eq 0$ for $i\in\{1,2\}$ and $0\leq j\leq 4$, we obtain
$\uc^{(1)}_i$ and hence
$\uc^{(2)}_i\eq \uc_i\xor\uc^{(1)}_i\xor\uc^{(0)}_i\in\C_2$,
$\C_2$ a $[15,5,11]$ RS code. In particular, $\uc^{(2)}_1$ has 5
errors and one erasure, so it is detected as uncorrectable, while
$\uc^{(2)}_2$ has four errors and two erasures, which
can be corrected. Once the errors and erasures in $\uc^{(2)}_2$ are
corrected, we obtain $\uc_2\eq
\uc^{(2)}_2\xor\uc^{(1)}_2\xor\uc^{(0)}_2$.

Finally, since $c_{i,j}\eq c^{(2)}_{i,j}$ for $0\leq i\leq 3$ and
$0\leq j\leq 4$, taking
the vectors $(c^{(2)}_{i,0},c^{(2)}_{i,1},\ldots,c^{(2)}_{i,4})\in (GF(16))^5$
for $i\in\{0,2,3\}$ as elements of a codeword in the (vertical) code
$\cV_0$, and since $\cV_0$ is a $[4,3,2]$ parity code over $(GF(16))^5$,
we retrieve $(c^{(2)}_{1,0},c^{(2)}_{1,1},\ldots,c^{(2)}_{1,4})$.
Encoding systematically this vector in $\C_2$, we obtain $\uc^{(2)}_1$, and
hence, $\uc_1\eq\uc^{(2)}_1\xor\uc^{(1)}_1\xor\uc^{(0)}_1$,
completing the decoding.
}
\end{example}

Theorem~\ref{theo3} gives a decoding algorithm for EII codes under
the assumption
that when the error-erasure correcting capability of a row is
exceeded, then the code successfully detects this situation and it
does not miscorrect. But what happens when there are miscorrections?
Error-erasure correcting algorithms of II-RS codes~\cite{cw,tk,w} and of EII-RS
codes~\cite{bh2} assume no miscorrections as well. Since these codes are
based on RS codes, the assumption boils down to the probability of
miscorrection of RS codes~\cite{b,kmc,kmc2,msw}. For example,
in~\cite{w} it is suggested that code $\C_0$ have at least 10 parity
symbols, which gives a reasonable low probability of miscorrection.
However, miscorrection may be relatively frequent in many cases,
specially when the codes $\C_i$ are not RS codes. Take for instance
the 3-level binary EII-code $\C(15,(4,8,10,15))$
according to
Definition~\ref{defEII}, where $\C_0$ is a $[15,11,3]$ Hamming code,
$\C_1\subset\C_0$ is a $[15,7,5]$ BCH code, $\C_2\subset\C_1$ is a
$[15,5,7]$ BCH code, $\cV_2$ is a (vertical)
$[4,1,4]$ repetition code over $(GF(2))^4$, $\cV_1$ is a $[4,2,3]$
extended RS code over $GF(4)$ and $\cV_0$ is a $[4,3,2]$ parity code over
$(GF(2))^5$. 
Assuming no erasures, since $\C_0$ is a perfect code, each time
more than one error occurs when attempting to decode a row, we will
have a miscorrection. Adapting the decoding algorithm of
Theorem~\ref{theo3} to include miscorrections may lead us to incorrect
decoding. Let us illustrate this point with a simple example.

\begin{example}
\label{ex16}
{\em
Consider a 2-level II-code $\C(7,(2,4))$ over $GF(8)$, where $\al$ is
a primitive element in $GF(8)$ such that\\ $1+\al+\al^3\eq 0$, $\C_0$
is a $[7,5,3]$ RS code generated by $(x+1)(x+\al)$, $\C_1$ is a
$[7,3,5]$ RS code generated by\\ $(x+1)(x+\al)(x+\al^2)(x+\al^3)$,
$\cV_0$ is a $[2,2,1]$ code over $(GF(8))^3$
(i.e., the whole space) and $\cV_1$ is a $[2,1,2]$ repetition code
over $(GF(8))^2$. According to Theorem~\ref{theo1}, $\C(7,(2,4))$ has
minimum distance 5.



Next, assume that the zero array is transmitted and
the following array with errors is received:

$$
\begin{array}{|c|c|c|c|c|c|c|}
\hline
0&0&0&0&\al^6&0&0\\
\hline
0&0&0&0&0&\al^2&\al^6\\
\hline
\end{array}
$$

Applying the error-erasure decoding algorithm of Theorem~\ref{theo3},
we first attempt to correct the rows in $\C_0$. Both are
correctable, giving the array

$$
\begin{array}{|c|c|c|c|c|c|c|}
\hline
0&0&0&0&0&0&0\\
\hline
0&0&0&0&1&\al^2&\al^6\\
\hline
\end{array}
$$

But this array has weight 3 and the minimum weight of $\C(7,(2,4))$
is 5, so it will be detected as an incorrect decoding. A possibility
is to apply the decoding algorithm by assuming that only one of the two
rows has been miscorrected. Assume that the first row was
correctly decoded, so after the first pass of the algorithm, by
leaving the second row unchanged, we have

$$
\begin{array}{|c|c|c|c|c|c|c|}
\hline
0&0&0&0&0&0&0\\
\hline
0&0&0&0&0&\al^2&\al^6\\
\hline
\end{array}
$$

Continuing the decoding algorithm, this array will be correctly
decoded as the zero array. On the other hand, if we assume
that the second row is the one correctly decoded, leaving the first
row intact, we have the following array:

$$
\begin{array}{|c|c|c|c|c|c|c|}
\hline
0&0&0&0&\al^6&0&0\\
\hline
0&0&0&0&1&\al^2&\al^6\\
\hline
\end{array}
$$

In this case the algorithm will produce the array

\begin{eqnarray*}
\begin{array}{|c|c|c|c|c|c|c|}
\hline
\al^4&0&0&\al^3&\al^6&0&0\\
\hline
0&0&0&0&1&\al^2&\al^6\\
\hline
\end{array}
\end{eqnarray*}
which we can verify that is a valid array in the code.
Absent any other assumptions, the
two solutions are equally likely.

However, suppose
that the zero array is transmitted and
the following array is received:

$$
\begin{array}{|c|c|c|c|c|c|c|}
\hline
0&0&0&0&\al^3&0&0\\
\hline
0&0&0&0&0&\al^2&\al^6\\
\hline
\end{array}
$$

In this case, again, if we assume that the first row has been
correctly decoded as the zero vector and the second row has more than
one error, then the decoding algorithm gives the zero array.
If we assume that the second row is the one that has been correctly
decoded and the first row has more than one error, then the decoding
algorithm will give an uncorrectable error when attempting to correct
two errors in the first row, hence, the solution is unique.
}
\end{example}

Example~\ref{ex16} shows that when we have miscorrections, we can
adapt the decoding algorithm to
different possibilities, but the solution may not be unique (in which
case we would be doing list decoding). The process is simple for
small values of $m$ and of $t$, as is the case in Example~\ref{ex16},
but it may be prohibitively
complicated for large $m$ and $t$. Of course, if there is a total of $x$
errors and $y$ erasures, where $2x+y\leq d-1$, $d$ the minimum
distance of $\C(n,\uu)$, the solution will be unique. We believe that
the subject of decoding errors and erasures in EII codes deserves
further research.

The following theorem
gives the erasure correcting capability of a $t$-level EII code.

\begin{theorem}
\label{theo2}
{\em
Consider the $t$-level EII code $\C(n,\uu)$ as given by
Definition~\ref{defEII}. Let $C\in\C(n,\uu)$ and assume that $C$ has erasures. Then $\C(n,\uu)$ can
correct in $C$ any row with up to $d^{\rm (H)}_{0}-1$ erasures, any up
to $d^{\rm (V)}_{i}-d^{\rm (V)}_{i-1}$ rows
with up to $d^{\rm (H)}_{t-i}-1$ erasures each for $1\leq i\leq t-1$
and any up to $d^{\rm (V)}_{0}-1$ rows with at least $d^{\rm
(H)}_{t-1}$ erasures each.
}
\end{theorem}
\pf Simply make $x_i\eq 0$ in Theorem~\ref{theo3}. Since there are no
errors, there cannot be miscorrection of a codeword. \qed

The next corollary completes the proof of Theorem~\ref{theo1}.

\begin{corollary}
\label {cor6}
{\em
Consider a $t$-level EII code $\C(n,\uu)$ as given by
Definition~\ref{defEII} and let $d$ be its minimum distance. Then
$$d\geq \min\{d_i^{\rm (H)}d_{t-1-i}^{\rm (V)}\,:\,0\leq i\leq t-1\}.$$
}
\end{corollary}

\noindent\pf Assume that there is a codeword $C\in\C(n,\uu)$ with $w$ erasures, where
$w\leq \min\{d_i^{\rm (H)}d_{t-1-i}^{\rm (V)}\,:\,0\leq i\leq
t-1\}-1$. We have to prove that such erasures can be corrected.

According to Theorem~\ref{theo2}, we may assume that the rows in $C$
containing erasures have at least $d_0^{\rm (H)}$ erasures each,
otherwise such rows get corrected. Also by Theorem~\ref{theo2}, it
suffices to prove that there up
to $d^{\rm (V)}_{i}-d^{\rm (V)}_{i-1}$ rows
with up to $d^{\rm (H)}_{t-i}-1$ erasures each for $1\leq i\leq t-1$
and up to $d^{\rm (V)}_{0}-1$ rows with at least $d^{\rm
(H)}_{t-1}$ erasures each in $C$. Assume that this claim is not true.
Then, there is an $i$, $0\leq i\leq t-1$, such that there are
at least $d^{\rm (V)}_{i}$ rows
with at least $d^{\rm (H)}_{t-1-i}$ erasures each in $C$. But if this
is the case, $C$ has at least $d_{t-1-i}^{\rm (H)}d_i^{\rm (V)}>w$ erasures, a
contradiction.
\qed

\section{Properties of EII-PC Codes}
\label{subpropEIIPC}
The next theorem gives necessary and sufficient conditions for
$t$-level EII-PC codes that will be useful to prove further properties.

\begin{theorem}
\label{theo4}
{\em
Consider a $t$-level EII-PC code
$\C(n,\uu)$ as given by
Definition~\ref{defEIIPC} and an $m\times n$ array $C$ with rows
$\uc_w$ for $0\leq w\leq m-1$. Then,
$C\in\C(n,\uu)$ if and only if $\uc_w\in\C_0$ for $0\leq w\leq m-1$
and, assuming $\C_t\eq\{0\}$,

\begin{eqnarray}
\label{eqEII-PC}
\bigoplus_{w=0}^{m-1}h_{r,w}\,\uc_w&\in& \C_{i}\;\;{\rm for}\;\; 1\leq
i\leq t\;\; {\rm and}\;\; 0\leq r\leq \hs_{i}-1.
\end{eqnarray}
}
\end{theorem}
\noindent\pf Let $C$ be an array in a $t$-level EII-PC code $\C(n,\uu)$ as given by
Definition~\ref{defEIIPC}. Hence, $C\eq\bigoplus_{v=0}^{t-1}C^{(v)}$, where $C^{(v)}$
is in the 1-level EII-PC code
$\C(n,\uu^{(v)})$ for $0\leq v\leq t-1$ and $\uu^{(v)}$ is given
by~(\ref{equui}). In particular, the rows of $C^{(v)}$ are in
the (horizontal) code $\C_{v}$ and the columns of $C^{(v)}$ are
in the (vertical) code $\cV^{(q)}_{t-v-1}$.
Denoting the rows of $C^{(v)}$ by $\uc^{(v)}_w$ for $0\leq w\leq
m-1$, hence
$\uc_w\eq\bigoplus_{v=0}^{t-1}\uc^{(v)}_w$. Since $\C_v\subset\C_0$,
in particular, $\uc^{(v)}_w\in\C_0$, so $\uc_w\in\C_0$.

Take $i$ such that 
$1\leq i\leq t$ and $r$ such that $0\leq r\leq \hs_{i}-1$, then,

\vspace{-3mm}

\begin{eqnarray}
\label{eqEII-PC2}
\bigoplus_{w=0}^{m-1}h_{r,w}\,\uc_w&=&\bigoplus_{w=0}^{m-1}h_{r,w}\,\bigoplus_{v=0}^{t-1}\uc^{(v)}_w
\quad =\quad
\bigoplus_{v=0}^{t-1}\,\,\,\bigoplus_{w=0}^{m-1}h_{r,w}\,\uc^{(v)}_w.
\end{eqnarray}

\vspace{-3mm}

In particular, if $v\,<\,i$, then $\hs_{v+1}\geq\hs_{i}\,>\,r$ and since
$\cV^{(q)}_{t-v-1}$ is a (vertical) $[m,m-\hs_{v+1}]$ code whose
parity-check matrix $H_{\hs_{v+1}\,,\,m\,,\,0}$ is given by~(\ref{Hvws}), we have
$\bigoplus_{w=0}^{m-1}h_{r,w}\,\uc^{(v)}_w\eq 0$ for $v\,<\,i$.
Thus, (\ref{eqEII-PC2}) becomes

\vspace{-3mm}

\begin{eqnarray}
\label{eqEII-PC3}
\bigoplus_{w=0}^{m-1}h_{r,w}\,\uc_w&=&\bigoplus_{v=i}^{t-1}\,\,\,\bigoplus_{w=0}^{m-1}h_{r,w}\,\uc^{(v)}_w.
\end{eqnarray}

Since $\uc^{(v)}_w\in\C_{v}\subseteq\C_{i}$ for $1\leq i\leq v\leq t$,
(\ref{eqEII-PC}) follows from~(\ref{eqEII-PC3}).

Conversely, assume that an $m\times n$ array $C$ with rows
$\uc_w$ satisfies $\uc_w\in\C_0$ for $0\leq w\leq m-1$
and~(\ref{eqEII-PC}). We proceed by induction on $t$. If $t\eq 1$,
then each row in $C$ is in is the $[n,n-u_0]$ (horizontal) code $\C_0$.
By~(\ref{eqEII-PC}), since $\C_1\eq\{0\}$,
$\bigoplus_{w=0}^{m-1}h_{r,w}\,\uc_w\eq 0$ for $0\leq r\leq
\hs_{1}-1$. Thus, every column of $C$ is in the (vertical)
$[m,m-\hs_1]$ code $\cV^{(q)}_0$.
According to Definition~\ref{defEIIPC}, $C$ is in the
1-level EII-PC code $\C(n,\uu)$.

So, take $t>1$ and assume that the result is true for $t-1$.
Denote by $(c_{x,y})_{\substack{0\leq x\leq m-1\\ 0\leq y\leq n-1}}$ the
entries of $C$, and in particular, consider the entries $c_{x,y}$
such that $0\leq x\leq m-\hs_t-1$, $0\leq y\leq n-u_{t-1}-1$. Encode such
elements into an array $C^{(t-1)}$ in the 1-level EII-PC code
$\C(n,\uu^{(t-1)})$
as given by Definition~\ref{defEIIPC}.
Each column in $C^{(t-1)}$
is in the $[m,m-\hs_t]$ (vertical) code
$\cV^{(q)}_{0}$ by construction, and so is each column in $C$
by~(\ref{eqEII-PC}) for $i\eq t$ and $0\leq r\leq \hs_t-1$ since $\C_t\eq\{0\}$, so $C$ and
$C^{(t-1)}$ coincide in their first $n-u_{t-1}$ columns. Thus, taking
$C'\eq C\xor C^{(t-1)}$, the first $n-u_{t-1}$ columns of
$C'$ are zero. Denote by $\uc'_w$ the rows of $C'$ and by
$\uc^{(t-1)}_w$ the rows of $C^{(t-1)}$, where $0\leq
w\leq m-1$. Claim: each $\uc'_w\in\C_0$ and

\vspace{-3mm}

\begin{eqnarray}
\label{eqEII-PC'}
\bigoplus_{w=0}^{m-1}h_{r,w}\,\uc'_w&\in& \C_{i}\;\;{\rm for}\;\; 1\leq
i\leq t-1\;\; {\rm and}\;\; 0\leq r\leq \hs_{i}-1.
\end{eqnarray}

\vspace{-3mm}

Since $\uc_w\in\C_0$ and $\uc^{(t-1)}_w\in\C_{t-1}\subset\C_0$ by construction,
then $\uc'_w\eq\uc_w\xor\uc^{(t-1)}_w\in\C_0$ for $0\leq w\leq m-1$.
Next take $i$ such that $1\leq
i\leq t-1$ and $0\leq r\leq \hs_{i}-1$. Since $\C_{t-1}\subseteq
\C_{i}$, then, in particular,
$\bigoplus_{w=0}^{m-1}h_{r,w}\,\uc^{(t-1)}_w\in\C_{i}$. Also,
by~(\ref{eqEII-PC}), in particular
$\bigoplus_{w=0}^{m-1}h_{r,w}\,\uc_w\in\C_{i}$, so~(\ref{eqEII-PC'}) follows.

By induction, this means that $C'$ is in a $(t-1)$-level EII-PC code
$\C(n,\tilde{\uu})$ according to Definition~\ref{defEIIPC}, where

\vspace{-3mm}

\begin{eqnarray*}
\tilde{\uu}&\eq &(\overbrace{u_0,u_0,\ldots,u_0}^{s_0}\,,\,\overbrace{u_1,u_1,\ldots,u_1}^{s_1}\,,\,\ldots\,,\,
\overbrace{u_{t-2},u_{t-2},\ldots,u_{t-2}}^{s_{t-2}})\,,\,\overbrace{u_{t-1},u_{t-1},\ldots,u_{t-1}}^{s_{t-1}+s_t})
\end{eqnarray*}
with nested horizontal codes $\C_{i}$ and nested vertical codes
$\cV^{(q)}_{t-1-i}$ for $0\leq i\leq t-2$.
Thus, $C\in\C(n,\uu^{(t-1)})\xor\C(n,\tilde{\uu})$, where
$\uu^{(t-1)}$ is given by~(\ref{equui}).
By Definition~\ref{defEIIPC},
$\C(n,\tilde{\uu})\eq\bigoplus_{v=0}^{t-2}\,\C(n,\uu^{(v)})$,
so $C\in\bigoplus_{v=0}^{t-1}\,\C(n,\uu^{(v)})$.
Again by Definition~\ref{defEIIPC}, $C\in\C(n,\uu)$.
\qed

\vspace{-3mm}

\begin{corollary}
\label{cor1}
{\em
Consider a $t$-level EII-PC code
$\C(n,\uu)$ as given by
Definition~\ref{defEIIPC} and an $m\times n$ array $C$ with rows
$\uc_w$ for $0\leq w\leq m-1$. Then,
$C\in\C(n,\uu)$ if and only if $\uc_w\in\C_0$ for $0\leq w\leq m-1$
and, assuming $\hs_{t+1}\eq 0$,

\begin{eqnarray}
\label{eqEII-PCbis}
\bigoplus_{w=0}^{m-1}h_{r,w}\,\uc_w&\in& \C_{i}\;\;{\rm for}\;\; 1\leq
i\leq t\;\; {\rm and}\;\; \hs_{i+1}\leq r\leq \hs_{i}-1.
\end{eqnarray}
}
\end{corollary}

\noindent\pf
Assume that~(\ref{eqEII-PCbis}) holds. Take $r$ such that $0\leq r\leq
\hs_{i+1}-1$ for $1\leq i\leq t-1$. In particular, there is a
$j$, $1\leq i<j\leq t$,
such that $\hs_{j+1}\leq r\leq
\hs_{j}-1$, then, by~(\ref{eqEII-PCbis}),
$\bigoplus_{w=0}^{m-1}h_{r,w}\,\uc_w\in \C_{j}$. Since
$\C_{j}\subset\C_{i}$, (\ref{eqEII-PC}) follows.
\qed

\begin{example}
\label{ex21}
{\em
Take the binary 3-level II-PC code $\C(8,(1,1,1,1,4,4,4,7))$  of
Example~\ref{ex20}. According to Corollary~\ref{cor1}, an $8\times 8$
array $C$ with rows $\uc_w$, $0\leq w\leq 7$, belongs in
$\C(8,(1,1,1,1,4,4,4,7))$, if and only if each $\uc_w$ belongs in the
$[8,7,2]$ parity-check code $\C_0$ and, by~(\ref{eqEII-PCbis}), using
the matrix $H_{8,8,0}$ given by~(\ref{eqH}) in Example~\ref{ex20},

$$
\begin{array}{ccccccccccccccccc}
\uc_0&\xor&\uc_1&\xor&\uc_2&\xor&\uc_3&\xor&\uc_4&\xor&\uc_5&\xor&\uc_6&\xor&\uc_7&\in&\C_2\\
\uc_0&\xor&\uc_1&    &     & \xor&\uc_3&\xor&\uc_4&&&&&& &\in&\C_1\\
\uc_0&&&\xor&\uc_2&\xor&\uc_3&&&\xor&\uc_5&&&&&\in&\C_1\\
&&\uc_1&\xor&\uc_2&\xor&\uc_3&&&&&\xor&\uc_6&&&\in&\C_1\\
\end{array}
$$
}
\end{example}

The properties of $t$-level EII-MDS and EII-RS codes are inherited from those of
$t$-level EII-PC codes. 
In effect, Corollary~\ref{cor1} gives:

\begin{corollary}
\label{cor2}
{\em
Consider a $t$-level EII-RS code
$\C(n,\uu)$ as given by
Definition~\ref{defEIIPC} and an $m\times n$ array $C$ with rows
$\uc_w$ for $0\leq w\leq m-1$. Then,
$C\in\C(n,\uu)$ if and only if $\uc_w\in\C_0$ for $0\leq w\leq m-1$
and, assuming $\hs_{t+1}\eq 0$,

\begin{eqnarray}
\label{eqEII-RS}
\bigoplus_{w=0}^{m-1}\al^{rw}\,\uc_w&\in& \C_{i}\;\;{\rm for}\;\; 1\leq
i\leq t\;\; {\rm and}\;\; \hs_{i+1}\leq r\leq \hs_{i}-1.
\end{eqnarray}
}
\end{corollary}
\qed

\vspace{-3mm}

The conditions $\uc_w\in\C_0$ for $0\leq w\leq m-1$
and~(\ref{eqEII-RS}) are given as the definition
of $t$-level II-RS codes for $s_t\eq 0$ in~\cite{bh,tk,w}, and of
$t$-level EII-RS codes in~\cite{bh2}. As we have seen,
$t$-level EII-RS codes are a special case of the more general class of
$t$-level EII-PC codes. Let us point out that in~\cite{hapkt}, 2-level
II-RS codes are called II codes, while in~\cite{tk,w}, $t$-level
II-RS codes for $t> 2$ are called Generalized II (GII) codes. Most papers on II codes
follow this convention~\cite{hys,z2,z3}. The reasons for this
denomination are historical, but we prefer to refer to EII codes
(which include II codes) as $t$-level EII codes because the value $t$
is stated explicitly and also, there is no conceptual difference
between the cases $t\eq 2$ and $t>2$.

Denote by $A\otimes B$ the Kronecker product between
matrices $A$ and $B$~\cite{ms}. Explicitly, if
$A\eq (a_{i,j})_{\substack{0\leq i\leq m_0-1\\ 0\leq j\leq n_0-1}}$
and\\
$B\eq (b_{u,v})_{\substack{0\leq u\leq m_1-1\\ 0\leq v\leq n_1-1}}$, then
$A\otimes B$ is the $(m_0m_1)\times (n_0n_1)$ matrix
$(a_{i,j}B)_{\substack{0\leq i\leq m_0-1\\ 0\leq j\leq n_0-1}}$, where $cB$
denotes the $m_1\times n_1$ matrix consisting of multiplying each
element of $B$ by $c$.

The Kronecker product is also called
the tensor product in literature~\cite{hys,if,wolf}.
The following lemma is a direct consequence of the definition of the
Kronecker product.

\begin{lemma}
\label{l8}
{\em Let $A$ be an $m_0\times n_0$ matrix, $B$ an $m_1\times n_1$
matrix and $C$ an 
$n_0\times n_1$ matrix. Denote by $\uc$ the $n_0n_1$
vector obtained by reading row-wise the elements of $C$. Let
$\uc^{\rm T}$ be the transpose of $\uc$ and
$\uu$ the $(m_0m_1)\times 1$ vector $\uu\eq (A\otimes B)\uc^{\rm T}$.
Then,

\begin{eqnarray}
\label{eqKr}
u_{rm_1+v}&=&\bigoplus_{w=0}^{n_0-1}a_{r,w}\bigoplus_{j=0}^{n_1-1}b_{v,j}\,c_{w,j}\quad
{\rm for}\quad 0\leq r\leq m_0-1\;,\;0\leq v\leq m_1-1.
\end{eqnarray}
\qed
}
\end{lemma}

The next theorem provides a parity-check matrix for a $t$-level
EII-PC code. The proof is similar to the one for $t$-level EII-RS
codes~\cite{bh2}, but we prove it for the sake of completeness. 

\begin{theorem}
\label{theo5}
{\em
Consider a $t$-level EII-PC code
$\C(n,\uu)$ over $GF(q)$ as given by Definition~\ref{defEIIPC},
where $u_t\eq n$. Using Corollary~\ref{cor1}, it can be proven that a
parity-check matrix of $\C(n,\uu)$ is given by the
$\left(mu_0+ns_t+\sum_{i=1}^{t-1}s_i(u_i-u_0)\right)\times
(mn)$ matrix

\begin{eqnarray}
\label{eqHnuu}
H(n,\uu) &=&\left(
\begin{array}{rcl}
I_m&\otimes & H_{u_0\,,\,n\,,\,0}\\
H_{s_{1}\,,\,m\,,\,\hs_{2}}&\otimes & H_{(u_1-u_0)\,,\,n\,,\,u_0}\\
H_{s_{2}\,,\,m\,,\,\hs_{3}}&\otimes & H_{(u_{2}-u_0)\,,\,n\,,\,u_{0}}\\
\vdots &\vdots &\vdots\\
H_{s_{i}\,,\,m\,,\,\hs_{i+1}}&\otimes & H_{(u_{i}-u_0)\,,\,n\,,\,u_{0}}\\
\vdots &\vdots &\vdots\\
H_{s_{t-1}\,,\,m\,,\,\hs_t}&\otimes & H_{(u_{t-1}-u_0)\,,\,n\,,\,u_{0}}\\
H_{s_{t}\,,\,m\,,\,0}&\otimes &I_n\\
\end{array}
\right)
\end{eqnarray}
where $I_v$ is a $v\times v$ identity matrix and $H_{s\,,\,w\,,\,v}$ is given
by~(\ref{Hvws}). 
}
\end{theorem}
\noindent\pf
Consider an $m\times n$ array $C$ with rows $\uc_v$, $0\leq v\leq
m-1$, such that $\uc_v\eq (c_{v,0},c_{v,1},\ldots,c_{v,n-1})$.
Writing row-wise the entries of $C$, we obtain the vector $\uc\eq
(\uc_0\,,\,\uc_1\,,\,\ldots\,,\,\uc_{mn-1})$ of length $mn$.
We have to prove that $C\in\C(n,\uu)$ if and only if $H(n,\uu)\,\uc^T\eq\uzero$, where $\uc^T$
is the transpose of the vector $\uc$
and $\uzero$ is a zero vector of length
$mu_0+ns_t+\sum_{i=1}^{t-1}s_i(u_i-u_0)$. 
by Corollary~\ref{cor1}, it
suffices to prove that $\uc_w\in\C_0$ for $0\leq w\leq m-1$ and~(\ref{eqEII-PCbis}) holds if and
only if $H(n,\uu)\,\uc^T\eq\uzero$.

Since a parity-check matrix of $\C_0$ is given by
$H_{u_0\,,\,n\,,\,0}$, then
$\uc_v\eq (c_{v,0},c_{v,1},\ldots,c_{v,n-1})\in\C_0$
for $0\leq v\leq
m-1$ if and only if, for $0\leq r\leq u_0-1$,
$\bigoplus_{w=0}^{n-1}\,h_{r,w}c_{v,w}\eq 0$, if and only if
$(I_m\otimes H_{u_0\,,\,n\,,\,0})\,\uc^T\eq \uzero_{\,mu_0}$,
where ${\uzero}_{\,s}$ denotes a zero vector of length $s$.

Next take~(\ref{eqEII-PCbis}) with $1\leq i\leq t-1$, then
$\bigoplus_{w=0}^{m-1}h_{r,w}\,\uc_w\in \C_{i}$  for
$\hs_{i+1}\leq r\leq \hs_{i}-1$, if and only if, since the parity-check matrix of
$\C_{i}$ is $H_{u_{i}\,,\,n\,,\,0}$ as given by~(\ref{Hvws}) and
since $\uc_w\in\C_0$,
$\bigoplus_{j=0}^{n-1}\,h_{v,j}\,c_{w,j}\eq 0$
for $0\leq v\leq u_0-1$, 

\begin{eqnarray*}
\bigoplus_{j=0}^{n-1}\,h_{v,j}\bigoplus_{w=0}^{m-1}\,h_{r,w}\,c_{w,j}\eq
\bigoplus_{w=0}^{m-1}\,h_{r,w}\bigoplus_{j=0}^{n-1}\,h_{v,j}\,c_{w,j}\eq
0\; {\rm for}\; 1\leq i\leq t-1,\, \hs_{i+1}\leq r\leq
\hs_{i}-1\; {\rm and}\; u_0\leq v\leq u_{i}-1,
\end{eqnarray*}
if and only if, by~(\ref{eqKr}) in Lemma~\ref{l8}, $(H_{s_{i}\,,\,m\,,\,\hs_{i+1}}\otimes
H_{(u_{i}-u_0)\,,\,n\,,\,u_{0}})\uc^T\eq\uzero_{s_{i}(u_{i}-u_0)}$
for $1\leq i\leq t-1$.

Taking~(\ref{eqEII-PCbis}) with $i\eq t$,
$\bigoplus_{w=0}^{m-1}h_{r,w}\,\uc_w\in \C_{t}\eq\{0\}$  for
$0\leq r\leq \hs_{t}-1$, if and only if
$\bigoplus_{w=0}^{m-1}h_{r,w}c_{w,j}\eq 0$ for $0\leq r\leq
\hs_t-1$ and $0\leq j\leq n-1$, if and only if
$(H_{s_{t}\,,\,m\,,\,0}\otimes I_n)\,\uc^{T}\eq\uzero_{\,n\,s_t}$.
\qed

\begin{example}
\label{ex22}
{\em
Take again the binary 3-level II-PC code $\C(8,(1,1,1,1,4,4,4,7))$  of
Examples~\ref{ex20} and~\ref{ex21}. According to~(\ref{eqHnuu}), the
parity-check matrix of $\C(8,(1,1,1,1,4,4,4,7))$ is the $23\times 64$
matrix

\begin{eqnarray*}
H(8,(1,1,1,1,4,4,4,7))&=&
\left(
\begin{array}{rcl}
I_8&\otimes & H_{1\,,\,8\,,\,0}\\
H_{3\,,\,8\,,\,1}&\otimes & H_{3\,,\,8\,,\,1}\\
H_{1\,,\,8\,,\,0}&\otimes & H_{6\,,\,8\,,\,1}\\
\end{array}
\right),
\end{eqnarray*}
where $H_{s\,,\,w\,,\,v}$ is given by~(\ref{Hvws}) and $H_{8,8,0}$
by~(\ref{eqH}).
Since the number of parities in the II code $\C(8,(1,1,1,1,4,4,4,7))$
is 23, this is the rank of $H(8,(1,1,1,1,4,4,4,7))$.

Similarly, consider the binary 2-level EII-PC code $\C(8,(1,1,1,1,4,4,4,8))$. According to~(\ref{eqHnuu}), the
parity-check matrix of $\C(8,(1,1,1,1,4,4,4,8))$ is the $25\times 64$
matrix

\begin{eqnarray*}
H(8,(1,1,1,1,4,4,4,8))&=&
\left(
\begin{array}{rcl}
I_8&\otimes & H_{1\,,\,8\,,\,0}\\
H_{3\,,\,8\,,\,1}&\otimes & H_{3\,,\,8\,,\,1}\\
H_{1\,,\,8\,,\,0}&\otimes & I_8\\
\end{array}
\right).
\end{eqnarray*}

In this case, $H(8,(1,1,1,1,4,4,4,8))$ is a $25\times
64$ matrix, and the number of parities in the code is 24, so we can
eliminate a row in $H(8,(1,1,1,1,4,4,4,8))$ to make it a matrix of
rank 24. However, it may be convenient at the decoding to have more
parities than the rank of the matrix, as is the case with product
codes.
}
\end{example}

Lemma~\ref{l2} states that the set of transpose arrays of a 1-level
EII-PC code is also a 1-level EII-PC code.
The next theorem generalizes Lemma~\ref{l2} to
$t$-level EII-PC codes. We give it without proof since it was proven
in~\cite{bh2} (Theorem~18) for the special case of EII-RS codes and the
general case proceeds similarly.

\begin{theorem}
\label{theo6}
{\em
Consider a $t$-level EII-PC code
$\C(n,\uu)$ as given by Definition~\ref{defEIIPC}, and take the set
of $n\times m$ arrays corresponding to the transpose arrays of the $m\times
n$ arrays in $\C(n,\uu)$. Then, this set of $n\times m$ arrays
constitute a $t$-level EII-PC code $\C(m,\uu^{\bf  (V)})$  
with horizontal codes
$\{0\}\eq\cV^{(q)}_t\subset\cV^{(q)}_{t-1}\subset \cV^{(q)}_{t-2}\subset \ldots
\subset\cV^{(q)}_0$ and vertical codes
$\{0\}\eq\C_t\subset\C_{t-1}\subset \C_{t-2}\subset \ldots
\subset\C_0$ such that

\begin{eqnarray}
\label{equuT}
\uu^{\bf  (V)} &=&
\left(\overbrace{\hs_t,\hs_t,\ldots,\hs_t}^{u_t-u_{t-1}}\,,\,\overbrace{\hs_{t-1},\hs_{t-1},\ldots,\hs_{t-1}}^{u_{t-1}-u_{t-2}}\,,\,\ldots\,,\,
\overbrace{\hs_1,\hs_1,\ldots,\hs_1}^{u_1-u_0}\,,\,\overbrace{\hs_0,\hs_0,\ldots,\hs_0}^{u_0}\right).
\end{eqnarray}
\qed
}
\end{theorem}

\vspace{-3mm}

In particular, from Theorem~\ref{theo6},
the set of transpose arrays of a
$t$-level EII-MDS code is also a $t$-level EII-MDS code. 

When decoding a $t$-level EII-PC code, by Theorem~\ref{theo6}, we can
first apply the decoding algorithm to rows. If there are errors and
erasures left, then we can apply the decoding algorithm to columns,
and so on, until either all errors and erasures are corrected, or
uncorrectable patterns remain. This process generalizes the usual
iterative row-column decoding algorithm of product codes. Assuming
that we are correcting erasures only, if after this iterative
row-column decoding algorithm erasures remain, we may attempt to
correct them by using the parity-check matrix of the code. If this is
not possible, an uncorrectable error is declared. Let us illustrate
the concepts in the next example.

\begin{example}
\label{ex23}
{\em
Take again the binary 3-level II-PC code $\C(8,(1,1,1,1,4,4,4,7))$  of
Examples~\ref{ex20} and~\ref{ex21} and the code of transpose arrays,
which, by Theorem~\ref{theo6}, is a 3-level EII-PC code
$\C(8,(0,1,1,1,4,4,4,8))$. Notice that $\C(8,(1,1,1,1,4,4,4,7))$ is
an II-PC code while $\C(8,(0,1,1,1,4,4,4,8))$ is an EII-PC code that is
not an II-PC code.

Consider the following
$8\times 8$ array in $\C(8,(1,1,1,1,4,4,4,7))$ with erasures:

\begin{center}
\begin{tabular}{c|c|c|c|c|c|c|c|c|}
\multicolumn{1}{c}{\phantom{0}}&\multicolumn{1}{c}{\footnotesize
0}&\multicolumn{1}{c}{\footnotesize 1}&\multicolumn{1}{c}{\footnotesize 2}&
\multicolumn{1}{c}{\footnotesize 3}& \multicolumn{1}{c}{\footnotesize
4}&\multicolumn{1}{c}{\footnotesize 5}&\multicolumn{1}{c}{\footnotesize
6}&\multicolumn{1}{c}{\footnotesize 7}\\
\cline{2-9}
{\footnotesize 0}&&E&&&&&&\\
\cline{2-9}
{\footnotesize 1}&E&&E&E&&&&E\\
\cline{2-9}
{\footnotesize 2}&&&&&&E&&\\
\cline{2-9}
{\footnotesize 3}&&&E&E&E&E&&\\
\cline{2-9}
{\footnotesize 4}&&&&E&&E&E&\\
\cline{2-9}
{\footnotesize 5}&&&E&E&&E&E&\\
\cline{2-9}
{\footnotesize 6}&&&&E&&&E&\\
\cline{2-9}
{\footnotesize 7}&&&&&&&&E\\
\cline{2-9}
\end{tabular}
\end{center}

Since, by Theorem~\ref{theo2}, $\C(8,(1,1,1,1,4,4,4,7))$ can correct
any row with up to one erasure, up to 3 rows with up to 3 erasures
each and up to one row with 7 erasures, the array cannot be corrected
by the erasure decoding algorithm in Theorem~\ref{theo2}.
Similarly,  since the code on
transpose arrays $\C(8,(0,1,1,1,4,4,4,8))$ requires that at least one
column is erasure-free, the pattern cannot be corrected by this code
either.
However, since each row
is in the $[8,7,2]$ code $\C_0$, rows 0, 2 and 7 can be corrected, 
and once this erasures are corrected, the remaining erasures
can be corrected by the code on transpose arrays
$\C(8,(0,1,1,1,4,4,4,8))$, showing the power of the
iterative decoding.
}
\end{example}

Example~\ref{ex2} compared a 1-level EII code with a 1-level EII-PC
of the same rate. Although the 1-level EII code had larger minimum
distance, the row-column iterative decoding algorithm allowed it to
correct more erasures on average. The next example shows that the
same may occur for $t$-level EII and EII-PC codes.

\begin{example}
\label{ex2bis}
{\em
Consider the binary 2-level EII code
$\C(16,(\overbrace{1,1,\ldots,1}^{5},\overbrace{4,4,\ldots,4}^{6}),\overbrace{16,16,\ldots,16}^{5}))$
as given by Definition~\ref{defEII}, where $\C_1\subset\C_0$,
$\C_0$ is a $[16,15,2]$ parity code,
$\C_1$ is a $[16,11,4]$ extended Hamming code,
$\cV^{(q)}_0$ is a $[16,11,6]$ code over $(GF(2))^{11}$ and
$\cV^{(q)}_1$ is a $[16,5,12]$ extended RS code over $GF(16)$. Let us
call this code $\C^{(a)}$. By
Theorem~\ref{theo1}, $\C^{(a)}$ is a $[256,141,d]$ binary code with
$d\geq 24$.

Consider next the binary 2-level EII-PC code
$\C(16,(\overbrace{1,1,\ldots,1}^{5},\overbrace{4,4,\ldots,4}^{6}),\overbrace{16,16,\ldots,16}^{5}))$
as given by Definition~\ref{defEIIPC}, where
$\C_0$ and $\C_1$ are like in $\C^{(a)}$, but $\cV^{(q)}_1\subset\cV^{(q)}_0$,
$\cV^{(q)}_0\eq\C_1$ and
$\cV^{(q)}_1$ is a $[16,5,8]$ binary extended BCH code. Let us call
this code $\C^{(b)}$. By
Corollary~\ref{cor6}, $d\geq 16$, but since there are arrays of
weight 16, $d\eq 16$ and
$\C^{(b)}$ is a $[256,141,16]$ binary code. Hence, $\C^{(a)}$ and $\C^{(b)}$
have the same length and dimension but $\C^{(b)}$ has smaller
minimum distance. However, a Montecarlo simulation gives that, if
erasures occur one after the other until an uncorrectable pattern is
obtained like in Example~\ref{ex2}, the EII code can correct on
average 38 erasures, while, applying the iterative decoding algorithm
to $\C^{(b)}$ and to its code of transpose arrays, $\C^{(b)}$ can correct on average 48
erasures.
}
\end{example}

\section{Uniform Distribution of the Parity Symbols}
\label{Uniform}

Given an $[mn,k]$
code over a field $GF(q)$ consisting of $m\times n$ arrays, if $mn-k\eq cm+r$, where
$0\leq r<m$, we say that the code has a balanced distribution of
parity symbols if there is a systematic encoding of the $k$ data
symbols into an $m\times n$ array such that $m-r$ of the rows contain
$c$ parity symbols, while
the remaining $r$ rows contain $c+1$ parity symbols.
Codes somewhat similar to II-RS codes with a balanced distribution of parity symbols
were presented in~\cite{chpb} for $r\eq 0$, i.e., $m$ divides $mn-k$ and
hence each row contains the same number $c$ of parity symbols.

Given a $t$-level EII-PC code $\C(n,\uu)$, we have so far placed the
symbols like in systematic Encoding Algorithm~\ref{algencodingsyst},
that is, at the end of each row in
non-decreasing order of the $u_i$s. However, this distribution of
symbols in general is not balanced.
If it can be shown that there is an uniform distribution of
erasures that can be corrected
by the code $\C(m,\uu^{\bf  (V)})$ (i.e., the code on transpose arrays as given by
Theorem~\ref{theo6}), then we can use those erasures as the locations
for the parity symbols.

We say that given a codeword $\uc$ in a code $\C$, $\uc$ has a burst
of erasures of length $\ell$ if exactly $\ell$ consecutive symbols in
$\uc$ are erased, including all-around cases.
The following theorem shows that,
using Theorem~\ref{theo6}, we
can obtain a balanced distribution of the parity symbols for a
$t$-level EII-PC code $\C(n,\uu)$ under certain conditions. 

\begin{theorem}
\label{theo8}
{\em
Consider a $t$-level EII-PC code $\C(n,\uu)$ as given by
Definition~\ref{defEIIPC}, and let
$\C(m,\uu^{\bf (V)})$ be the $t$-level EII-PC code of transpose arrays as given by
Theorem~\ref{theo6}. Assume that each vertical code $\cV^{(q)}_i$ can
correct any burst of erasures of length $\hs_{t-i}$ for $0\leq i\leq t-1$.
Then $\C(n,\uu)$ has a balanced distribution of the parity symbols.
}
\end{theorem}

\noindent\pf We need to find $s\eq\sum_{i=0}^ts_iu_i$ erasures such that,
if $s\eq cm+r$ with $0\leq r<m$, then there are
$m-r$ rows with $c$ erasures each and $r$ rows with $c+1$
erasures each, and the erasures are correctable by the $t$-level
EII code $\C(m,\uu^{\bf (V)})$ on columns as given by
Theorem~\ref{theo6}. Then such erasures can be used to place the
parity symbols.

In effect, let $v_0\geq v_1\geq \ldots\geq v_{z-1}$ be the
non-zero elements of $\uu^{\bf (V)}$ in non-increasing order. In particular,
$s\eq \sum_{i=0}^{z-1}v_i$. We will
select the first $z$ columns in an $m\times n$ array such that column
$j$ has a burst of $v_j$ erasures for $0\leq j\leq z-1$. Then, by the
decoding algorithm given in
Theorem~\ref{theo3}, such erasures are correctable. In addition, we
will show that the selection of erasures is balanced. We proceed by
induction on $z$.

If $z\eq 1$, we have only one column that can correct a burst of
$v_0$ erasures and we place the erasures in the
top $v_0$ positions of that column. In particular, the distribution
is balanced (the top $v_0$ rows contain one erasure and the remaining
ones no erasures). So assume that $z>1$.

Consider the first $z-1$ columns and let $s'\eq \sum_{i=0}^{z-2}v_i$. By
induction, if $s'\eq c'm+r'$, we can place a burst of $v_j$ erasures in column
$j$ for $0\leq j\leq z-2$, such that the first $r'$ rows contain
$c'+1$ erasures and the last $m-r'$ rows contain $c'$ erasures.

If $v_{z-1}\,\leq\,m-r'$, then in column $z-1$ we place the $v_{z-1}$
erasures in locations $r',r'+1,\ldots ,r'+v_{z-1}-1$. Then the first
$r'+v_{z-1}$ rows contain $c'+1$ erasures and the last
$m-(r'+v_{z-1})$ rows contain $c'$ erasures, giving a balanced
distribution of the erasures.

If $v_{z-1}>m-r'$, then in column $z-1$ we place the $v_{z-1}$
erasures in locations $$0,1,\ldots ,v_{z-1}-(m-r')-1,
r',r'+1,\ldots ,m-1.$$ Then the first $v_{z-1}-(m-r')$ rows contain
$c'+1$ erasures and the
remaining rows $c'$ erasures, also giving a
balanced distribution of the erasures.
\qed

\begin{corollary}
\label{cor8}
{\em
Consider a $t$-level PC-EII code $\C(n,\uu)$ as given by
Definition~\ref{defEIIPC}, and consider the code
$\C(m,\uu^{\bf (V)})$ of transpose arrays as given by
Theorem~\ref{theo6}. Assume that each vertical code $\cV^{(q)}_i$ is a
cyclic code for $0\leq i\leq t-1$.
Then $\C(n,\uu)$ admits a balanced distribution
of the parity symbols.
}
\end{corollary}
\pf It suffices to prove that an $[n,k]$ cyclic code can correct any
burst of $n-k$ erasures and the result follows from
Theorem~\ref{theo8}. Since the code can encode systematically $k$
symbols into $n$ symbols, where the first $k$ symbols are the data
symbols, the encoding process can be viewed as the correction of
$n-k$ erasures in the last $n-k$ symbols. Any other codeword with a
burst of $n-k$ erasures, through a rotation, can be transformed into
a codeword with $n-k$ erasures in the last $n-k$ symbols since the
code is cyclic, so such $n-k$ consecutive erasures can be recovered.
\qed

The balanced distribution of parity symbols is certainly not unique.
We illustrate the method described
in Theorem~\ref{theo8} in the next example.

\begin{example}
\label{ex26}
{\em
Consider the binary 2-level II-PC code $\C(8,(1,1,1,1,7,7,7,7))$
where $\C_0$ is the $[8,7,2]$ parity code, $\C_1$ the $[8,1,8]$
repetition code, $\cV^{(2)}_0$ the $[8,8,1]$ code corresponding to the
whole space and $\cV^{(2)}_1$ a cyclic $[8,4,4]$ extended Hamming code.
By Theorem~\ref{theo6},
the code consisting of the transpose arrays is a
2-level EII-PC code $\C(8,(0,4,4,4,4,4,4,8))$ code. The uniform
distribution of parities given in the proof of Theorem~\ref{theo8} is
the following:

\begin{center}
\begin{tabular}{c|c|c|c|c|c|c|c|c|}
\multicolumn{1}{c}{\phantom{0}}&\multicolumn{1}{c}{\footnotesize
0}&\multicolumn{1}{c}{\footnotesize 1}&\multicolumn{1}{c}{\footnotesize 2}&
\multicolumn{1}{c}{\footnotesize 3}& \multicolumn{1}{c}{\footnotesize
4}&\multicolumn{1}{c}{\footnotesize 5}&\multicolumn{1}{c}{\footnotesize
6}&\multicolumn{1}{c}{\footnotesize 7}\\
\cline{2-9}
{\footnotesize 0}&E&E&&E&&E&&\\
\cline{2-9}
{\footnotesize 1}&E&E&&E&&E&&\\
\cline{2-9}
{\footnotesize 2}&E&E&&E&&E&&\\
\cline{2-9}
{\footnotesize 3}&E&E&&E&&E&&\\
\cline{2-9}
{\footnotesize 4}&E&&E&&E&&E&\\
\cline{2-9}
{\footnotesize 5}&E&&E&&E&&E&\\
\cline{2-9}
{\footnotesize 6}&E&&E&&E&&E&\\
\cline{2-9}
{\footnotesize 7}&E&&E&&E&&E&\\
\cline{2-9}
\end{tabular}
\end{center}

Since the codes $\cV^{(2)}_0$ and $\cV^{(2)}_1$ are cyclic, the erasures are correctable by
Corollary~\ref{cor8}.

Similarly, if we consider the EII-PC code $\C(8,(0,4,4,4,4,4,4,8))$,
the uniform distribution of parities given in the proof of
Theorem~\ref{theo8} is the following: 

\begin{center}
\begin{tabular}{c|c|c|c|c|c|c|c|c|}
\multicolumn{1}{c}{\phantom{0}}&\multicolumn{1}{c}{\footnotesize
0}&\multicolumn{1}{c}{\footnotesize 1}&\multicolumn{1}{c}{\footnotesize 2}&
\multicolumn{1}{c}{\footnotesize 3}& \multicolumn{1}{c}{\footnotesize
4}&\multicolumn{1}{c}{\footnotesize 5}&\multicolumn{1}{c}{\footnotesize
6}&\multicolumn{1}{c}{\footnotesize 7}\\
\cline{2-9}
{\footnotesize 0}&E&E&E&E&&&&\\
\cline{2-9}
{\footnotesize 1}&E&E&E&E&&&&\\
\cline{2-9}
{\footnotesize 2}&E&E&E&E&&&&\\
\cline{2-9}
{\footnotesize 3}&E&E&E&E&&&&\\
\cline{2-9}
{\footnotesize 4}&E&E&E&&E&&&\\
\cline{2-9}
{\footnotesize 5}&E&E&&E&&E&&\\
\cline{2-9}
{\footnotesize 6}&E&&E&E&&&E&\\
\cline{2-9}
{\footnotesize 7}&&E&E&E&&&&E\\
\cline{2-9}
\end{tabular}
\end{center}
}
\end{example}

The next corollary corresponds to Theorem~21 for EII-RS codes in~\cite{bh2}.

\begin{corollary}
\label{cor9}
{\em
Consider a $t$-level EII-MDS code $\C(n,\uu)$.
Then $\C(n,\uu)$ admits a balanced distribution
of the parity symbols.
}
\end{corollary}
\pf Consider the $t$-level EII-MDS code $\C(m,\uu^{\bf (V)})$ of
transpose arrays as given by
Theorem~\ref{theo6}. Each vertical code $\cV^{(q)}_i$ is an
$[m,m-\hs_{t-i},\hs_{t-i}+1]$ code for $0\leq i\leq
t-1$, so, in particular, it can correct a burst of erasures of length
$\hs_{t-i}$ and the result follows from Theorem~\ref{theo8}.
\qed


\section{Ordering of the Symbols of $t$-level EII-PC Codes Maximizing
Burst Correction}
\label{burst}
Consider the following problem: given a $t$-level EII-PC code
$\C(n,\uu)$, we want to map the codeword array into a sequence of
transmitted symbols in such a way that the burst-correcting
capability of the code is maximized. For simplicity, let us assume an
erasure only model. If the symbols are transmitted row-wise,
the burst-correcting capability is not maximized in general. The total
number of (independent) parities is $\sum_{i=0}^{m-1}s_iu_i$. Thus,
the maximum length of a burst that any ordering of the symbols can
correct is upper bounded by this number. Let us illustrate the
concept by taking a very simple example:

\begin{example}
\label{ex11}
{\em
Consider a $4\times 4$ product code with parity on rows and columns.
Consider the following two possible orderings of the symbols,
the regular row-wise ordering on the left, and the
diagonal ordering on the right:

$$
\begin{array}{cc}
\begin{array}{|c|c|c||c|}
\hline
0&1&2&3\\ \hline
4&5&{\textcolor{red}{6}}&{\textcolor{red}{7}}\\ \hline
{\textcolor{red}{8}}&{\textcolor{red}{9}}&{\textcolor{red}{10}}&11\\ \hline
\hline
12&13&14&15\\ \hline
\end{array}
&
\begin{array}{|c|c|c||c|}
\hline
0&4&{\textcolor{red}{8}}&{\textcolor{red}{12}}\\ \hline
13&1&5&{\textcolor{red}{9}}\\ \hline
{\textcolor{red}{10}}&14&2&{\textcolor{red}{6}}\\ \hline
\hline
{\textcolor{red}{7}}&{\textcolor{red}{11}}&15&3\\ \hline
\end{array}
\end{array}
$$

It is easy to see that the regular read-out can correct any burst of
length up to 5 (for example, the one starting in symbol 6 in red), but not
all the bursts of length 6, like the one starting in symbol 6, while
the diagonal read-out can correct any burst of length up to 7 (for
example, the one starting in symbol 6 in red). Since the number of
(independent) parity symbols is 7, the diagonal
ordering of the symbols meets the upper bound on the length of a
burst that the code can correct, i.e., it is optimal.



}
\end{example}

The product code in Example~\ref{ex11} is a
1-level EII-PC code $\C(4,(1,1,1,4))$. Let us consider next general
$t$-level EII-PC codes. For simplicity, we assume that the EII-PC codes are
EII-MDS codes. The problem of finding orderings of the symbols
optimizing burst correction for $t$-level II-MDS codes (that is,
$s_t\eq 0$) was studied in~\cite{bcl,mv}. These two references study
the burst error correcting problem, but for erasures the treatment is
the same.

Consider a $t$-level EII-MDS code
$\C(n,\uu)$ and let $\uu\eq (v_0,v_1,\ldots,v_{m-1})$. We say that $\C(n,\uu)$
is continuous if $v_{i+1}-v_i\leq 1$ for $0\leq i\leq
m-2$, and we say that it is symmetric if $v_i+v_{m-1-i}\eq v_j+v_{m-1-j}$
for $0\leq i,j\leq m-1$. For example, a $t$-level EII-MDS code
$\C(n,(1,1,2,3))$ is continuous (but not symmetric), while a
$t$-level EII-MDS code $\C(n,(1,1,2,3,3))$ is both continuous and
symmetric. In~\cite{bcl}, an optimal ordering of the symbols was
given for $t$-level II-MDS codes $\C(n,\uu)$ that are continuous and symmetric
provided that $n$ is a multiple of $v_0+v_{m-1}$ (the algorithm giving
the optimal ordering is given by example in~\cite{bcl} and it is formalized
in~\cite{mv}).

Given a $t$-level EII-MDS code $\C(n,\uu)$, 
let $\delta\eq\max\{1,
\max\{v_{i+1}-v_i\,:\,0\leq i\leq m-2\}\}$.
In~\cite{mv}, it
was proven that for any ordering of the symbols of a $t$-level II-MDS
code (hence, $s_t\eq 0$), the maximum length of a burst that the code can correct is at
most $(\sum_{i=0}^{t-1}s_iu_i)-\delta +1$. In particular, if $\delta\eq
1$, the code is continuous and the bound is given by the number of
parity symbols $\sum_{j=0}^{m-1}v_j\eq\sum_{i=0}^{t-1}s_iu_i$. Moreover, in~\cite{mv} an
algorithm giving an ordering that meets the bound is presented for
$t$-level II-MDS codes $\C(n,\uu)$ (i.e., $s_t\eq 0$) provided that $n$ is a multiple of
$\sum_{i=0}^{t-1}s_iu_i$. However, the bound is limited to the
decoding of the code $\C(n,\uu)$ only. The
bound in~\cite{mv} for II-MDS codes is valid for EII-MDS codes as
well. The question is, can the bound be improved when
applying the iterative row-column decoding algorithm using both
$\C(n,\uu)$ and $\C(m,\uu^{\bf (V)})$, where $\uu^{\bf  (V)}$ is
given by~(\ref{equuT})?

The answer is yes.
In effect, consider the $4\times 4$ product code with parity on
rows and columns described in Example~\ref{ex11}, which is a 1-level EII-MDS code
$\C(4,(1,1,1,4))$.
Since $\delta\eq 3$, the maximal length of a burst the code can
correct is 5. The procedure assumes that at each step, the code corrects
up to 3 rows with up to one erasure each and one row with up to 4
erasures. But
we have seen that applying row-column decoding to the diagonal ordering of
symbols given in Example~\ref{ex11}, we can correct any burst of
length up to 7, which is the number of parity bits.

Combining the result from~\cite{mv} and Theorem~\ref{theo6}, we have
the following lemma:

\begin{lemma}
\label{l4}
{\em Consider the $t$-level EII-MDS codes $\C(n,\uu)$ and
$\C(m,\uu^{\bf V})$ as given by Theorem~\ref{theo6}. Let $\uu\eq
(v_0,v_1,\ldots,v_{m-1})$, \\
$\uu^{\bf V}\eq (v'_0,v'_1,\ldots,v'_{n-1})$,
$\delta^{\bf H}\eq\max\{1,\max\{v_{i+1}-v_i\,:\,0\leq i\leq m-2\}\}$,
$\delta^{\bf V}\eq\max\{1,\max\{v'_{i+1}-v'_i\,:\,0\leq i\leq
n-2\}\}$ and $\delta\eq\min\{\delta^{\bf H},\delta^{\bf V}\}$.
Then, given any ordering of the symbols in the arrays,
the maximum length of a burst that can be corrected using either
$\C(n,\uu)$ or $\C(m,\uu^{\bf V})$ is
$(\sum_{i=0}^ts_iu_i)-\delta+1$. \qed
}
\end{lemma}

Let us take next an example with $t>1$.

\begin{example}
\label{ex12}
{\em
Consider a 2-level II-MDS code $\C(n,(1,3))$ over $GF(q)$, where $n<q$.
In this case, $\delta^{\bf H}\eq 2$, where $\delta^{\bf H}$ was
defined in Lemma~\ref{l4}, so the maximum length of a burst that any
ordering of the symbols of $\C(n,(1,3))$ can correct, according to
the bound in~\cite{mv}, is $(1+3)-\delta^{\bf H}+1\eq 3$. Moreover, as shown
in~\cite{mv}, there is an ordering achieving this bound for each $n$
such that $n$ is a multiple of 4. For example, if we take 
the usual row-wise ordering, we can correct any burst of length 3,
since such a burst would either have 3 erasures in the same row or
one erasure in one row and two in the other one.



Let $n\eq 4$. By Theorem~\ref{theo6}, the 2-level EII-MDS code of
transpose arrays of $\C(4,(1,3))$
is the 2-level EII-MDS code $\C(2,(0,1,1,2))$. This
code is continuous and symmetric. According to the
result in~\cite{bcl}, there is an ordering correcting any burst of
length 4. Specifically, consider the following ordering:



$$
\begin{array}{|c|c|c|c|}
\hline
0&{\textcolor{red}{1}}&{\textcolor{red}{3}}&7\\ \hline
{\textcolor{red}{2}}&{\textcolor{red}{4}}&5&6\\ \hline
\end{array}
$$

Notice that any burst of length 4 consists of 3 erased symbols in one of the
rows and the remaining erased symbol in the next row, with the exception of
the burst involving symbols 1,2,3,4 (in red).
The (horizontal) code $\C(4,(1,3))$ cannot correct this burst using
the decoding algorithm of Theorem~\ref{theo2}, since it
contains two erased symbols in each row. However, we can easily
verify that all the bursts of
length 4, including all-around bursts, contain two columns with one
erasure and one column with 2 erasures, while a fourth column is
erasure-free  (in particular, the burst in
red above). Hence, all
the bursts of length 4 with the above ordering can be corrected by
the (vertical) code $\C(2,(0,1,1,2))$ using the decoding algorithm of
Theorem~\ref{theo2}.

}
\end{example}

Example~\ref{ex12} shows that Lemma~\ref{l4} may improve the bound
in~\cite{mv} by using either the code $\C(n,\uu)$ or the
code of transpose arrays $\C(m,\uu^{\bf V})$. The next example shows
that the bound in Lemma~\ref{l4} can be improved even further
with iterative decoding using both codes.

\begin{example}
\label{ex13}
{\em
Consider a 2-level II-MDS code $\C(4,(1,1,2,4))$ over $GF(4)$.
In this case, the code of transpose arrays, by Theorem~\ref{theo6}, is also
a 3-level II-MDS code $\C(4,(1,1,2,4))$ and $\C_0\eq\cV_0$
is a $[4,3,2]$ code while $\C_1\eq\cV_1$ is a $[4,2,3]$ code, both
codes extended RS codes over $GF(4)$. Since
$\delta\eq\delta_1\eq\delta_2\eq 2$, by Lemma~\ref{l4}, if we
decode using either $\C(4,(1,1,2,4))$ or the code of transpose
arrays, the maximum
length of a burst that can be corrected is 7. However, consider the
following ordering of the symbols with the bursts of length 8
starting in 0, 8 and 12 (all-around burst) respectively in red:

$$
\begin{array}{ccc}
\begin{array}{|c|c|c|c|}
\hline
{\textcolor{red}{0}}&{\textcolor{red}{1}}&{\textcolor{red}{3}}&10\\\hline
9&{\textcolor{red}{2}}&{\textcolor{red}{7}}&11\\\hline
14&13&{\textcolor{red}{5}}&{\textcolor{red}{4}}\\\hline
8&{\textcolor{red}{6}}&15&12\\\hline
\end{array}
&
\begin{array}{|c|c|c|c|}
\hline
0&1&3&{\textcolor{red}{10}}\\\hline
{\textcolor{red}{9}}&2&7&{\textcolor{red}{11}}\\\hline
{\textcolor{red}{14}}&{\textcolor{red}{13}}&5&4\\\hline
{\textcolor{red}{8}}&6&{\textcolor{red}{15}}&{\textcolor{red}{12}}\\\hline
\end{array}
&
\begin{array}{|c|c|c|c|}
\hline
{\textcolor{red}{0}}&{\textcolor{red}{1}}&{\textcolor{red}{3}}&10\\\hline
9&{\textcolor{red}{2}}&7&11\\\hline
{\textcolor{red}{14}}&{\textcolor{red}{13}}&5&4\\\hline
8&6&{\textcolor{red}{15}}&{\textcolor{red}{12}}\\\hline
\end{array}
\end{array}
$$

We can see that these three bursts of length 8 can be corrected using
the iterative decoding using both codes, and the reader can verify
that the same is true for any of the 16 bursts of length 8, including
all-around bursts. Since the number of parities of the code is 8,
this ordering achieves the upper bound on the maximum
burst-correcting capability.
}
\end{example}

Future research requires determining if the upper bound
$\sum_{i=0}^{t}s_iu_i$ for the maximum length of a correctable
burst using the iterative decoding algorithm on codes $\C(n,\uu)$ and
$\C(m,\uu^{\bf V})$ can always be achieved (as in
Example~\ref{ex13}), and if not, finding
what such maximum correctable burst length is.

\section{Conclusions}
\label{conclusions}
We have presented a general approach to Integrated Interleaved and
Extended Integrated Interleaved codes. More traditional approaches
involve Reed-Solomon types of codes over a field $GF(q)$ such that
$q$ has size at least the length of the rows and columns in the
array. Our new general approach involves describing a $t$-level EII
code as a direct sum of 1-level EII codes. This approach allows for a
generalization of EII codes to any field. An important special case
involves taking product codes as 1-level EII codes. This special case
allows for iterative decoding on rows as well as on columns,
generalizing product codes. We discussed encoding, decoding, and
several applications, like the ordering of the symbols in the code
optimizing its burst correcting capability.

\end{document}